\documentclass[twocolumn]{aastex7}
\usepackage{amsmath}

\newcommand{\ta}[0]{\tablenotemark{a}}
\newcommand{\tb}[0]{\tablenotemark{b}}
\newcommand{\tc}[0]{\tablenotemark{c}}

\newcommand{\dpn}[3]{$#1\,{\rm #2}_{#3}$}

\begin{document}

\title{The Effect of Massive Trans-Neptunian Objects in the Long-term Evolution and Leaking Rates of Neptune's 3:2 and 2:1 Mean Motion Resonances}

\author[orcid=0000-0002-0792-4332,gname=Marco, sname=Mu\~noz-Guti\'errez]{Marco A. Mu\~noz-Guti\'errez}
\affiliation{Instituto de Astronom\'ia y Ciencias Planetarias, Universidad de Atacama, Copayapu 485, Copiap\'o, Chile}
\email[show]{marco.munoz@uda.cl}

\correspondingauthor{Marco A. Mu\~noz-Guti\'errez}
%\email[show]{marco.munoz@uda.cl}

\author[orcid=0009-0007-9542-0444,gname=Sebasti\'an, sname=Ram\'irez]{Sebasti\'an Ram\'irez-Vargas}
\affiliation{Instituto de Astrof\'isica, Pontificia Universidad Cat\'olica de Chile, Av. Vicu\~na Mackenna 4860, 782-0436 Macul, Santiago, Chile}
\email{sramrev@uc.cl}

\author[orcid=0000-0001-7042-2207,gname=Antonio, sname=Peimbert]{Antonio Peimbert}
\affiliation{Instituto de Astronom\'ia, Universidad Nacional Aut\'onoma de M\'exico, Apdo. postal 70-264, Ciudad Universitaria, M\'exico}
\email{antonio@astro.unam.mx}

\author[orcid=0000-0002-5974-3998,gname=Angeles, sname=P\'erez-Villegas]{Angeles P\'erez-Villegas}
\affiliation{Instituto de Astronom\'ia, Universidad Nacional Aut\'onoma de M\'exico, A. P. 106, C.P. 22800, Ensenada, B.C., M\'exico}
\email{mperez@astro.unam.mx}

\author[orcid=0000-0003-0412-9314,gname=Cristobal, sname=Petrovich]{Cristobal Petrovich}
\affiliation{Millennium Institute for Astrophysics, Santiago, Chile}
\affiliation{Department of Astronomy, Indiana University, Bloomington, IN 47405, USA}
\email{cpetrovi@iu.edu}


\begin{abstract}

The current populations trapped in Neptune's main mean motion resonances in the Kuiper belt, Plutinos in the 3:2 and Twotinos in the 2:1, contain some of the best-characterized minor objects in the Solar System, given their dynamical importance. In particular, Twotinos may hide evidence of Neptune's early migration. However, these populations vary in time, declining at a rate that has not been previously clearly established. In this work, we use numerical simulations to study the long-term evolution of the Plutino and Twotino populations. We use two data sources: the most up-to-date observations and the theoretical debiased model of the Kuiper belt known as L7. In addition to studying the giant planets' effect on these populations over 4 Gyr, we analyze the additional impact produced by the ten most massive trans-Neptunian objects (TNOs) trapped in these resonances, as well as the effect of Pluto on the 2:1 population. We find that the decay rate in each resonance can be modeled as a stochastic process well described by an exponential decay with an offset determined by an underlying long-term stable population. The most massive TNOs, particularly Pluto, influence this decay rate significantly, as expected for the 3:2 resonance. Remarkably, Pluto also strongly influences the 2:1 resonance's evolution.

\end{abstract}




\section{Introduction} 
\label{sec:intro}

Neptune's mean motion resonances (MMRs) in the Kuiper belt (KB) are dynamical regions where a significant fraction of the total observed trans-Neptunian objects (TNOs) are located \citep[e.g.][]{Lykawka07,Gladman12,Alexandersen16,Volk16}. For example, in the full data release of the Origin of the Solar System Objects Survey \citep[OSSOS;][]{Bannister18}, 132 Plutinos are listed out of a total of 838 well-characterized TNOs\footnote{From a total of 840 TNOs discovered in the OSSOS survey, only two objects were not tracked to high-precision orbits.}, while another 34 in the 2:1 \citep[a.k.a. Twotinos;][]{Chen19}, 39 objects in the 7:4, and 29 objects in the 5:2 MMRs \citep{Crompvoets22} were found in the survey; those numbers show that $\sim$27\% of the total OSSOS-characterized discoveries were located in just four Neptune MMRs. More recently, \citet{Forgacs23} performed a classification of Kuiper belt objects (KBOs) located in MMRs, finding that from the 4121 objects with semimajor axis $a>30.1$ au, listed in the JPL Horizons database\footnote{\url{https://ssd.jpl.nasa.gov/horizons/}}, 906 librate, for at least 100~Myr, in one of several possible MMRs with Neptune; their resonant objects were identified using their custom Fast Identification of MMRs method \citep[or FAIR method,][]{Forgacs18}. Based on these results, $\sim22\%$ of the reported TNOs, observed by various surveys, are located within the MMRs with Neptune. 

Two of the most significant resonances associated with Neptune are the 3:2 and 2:1. This is important for several reasons. First, most well-characterized resonant objects belong to the 3:2 population, often referred to as ``Plutinos''. This group includes Pluto, the most notable and well-studied trans-Neptunian object (TNO), both physically and dynamically \citep[e.g.][]{Milani89,Stern18}. Additionally, the locations of the 3:2 and 2:1 resonances define the inner and outer boundaries of the main classical Kuiper Belt, respectively \citep[see][]{Gladman08}. Moreover, the orbital distribution of the 2:1 resonance population, known as ``Twotinos'', could provide crucial evidence for understanding the characteristics of Neptune's early migration \citep{Ida00,Chiang02,Wyatt03,Murray05,Chen19,Li23}.

Indeed, the populations of Neptune's MMRs are thought to originate from the very early stages of solar system evolution, during the giant planets' outward migration stage, due to the interaction of the giants with a massive disk of remaining planetesimals \citep{Fernandez84,Nesvorny18}. The first theory to explain the population of MMRs (and the origin of Pluto's orbit in particular) is due to \citet{Malhotra93,Malhotra95}. In this framework, often referred to as ``adiabatic resonance sweeping", the slow, smooth migration of Neptune carries its first-order MMRs through an external, dynamically cold planetesimal belt (including Pluto). Planetesimals captured into these resonances are transported outward along with Neptune, while their eccentricities—and, to a lesser extent, their inclinations—are adiabatically excited \citep[e.g.][]{Malhotra98,Gomes00}. 

While this process sweeps an overly large number of objects into the MMRs, their eccentricities naturally increase, possibly bringing them close enough to Neptune or Uranus to eject them from the KB \citep[e.g.][]{Morbidelli97,Munoz18}. Recently, a non-smooth, but instead grainy migration of Neptune has been favoured to explain the relatively small number of particles trapped in the resonances \citep{Nesvorny16,Lawler19}, as sudden jumps during the migration would release some of the trapped TNOs, making such a migration less efficient in trapping objects into the MMR.

An additional possibility studied in the literature is the existence of a rogue planet present in the outer solar system during its early stages; such an object can enhance the capture efficiency of Neptune's far MMRs \citep{Huang22}. Although this scenario has only been studied farther out than the 2:1 MMRs, its efficiency could be of importance for closer resonances as well.

Once the planetary migration ends, 
a few to several dozens of Myr after the formation of the Solar System, it is believed that the populations trapped in MMRs will evolve steadily, provided escape rates are low; this would imply, at the same time, that ratios between population sizes of different MMRs should only slowly evolve along the solar system's age. \citep[e.g.][]{Hahn05,Tiscareno09}.

On the other hand, it is well known that a leaking process from MMRs has constantly occurred over the age of the Solar System, either due to a weak chaotic diffusion present inside MMRs or by the direct perturbations of massive bodies inside and outside of the resonances \citep{Morbidelli97,Ip97,Tiscareno09,Munoz19}. In contrast to the above process, a refilling or resupplying of those same MMRs with new material, mainly coming from the classical KB, has not been studied nor considered important when estimating the long-term evolution of the populations trapped inside MMRs with Neptune. Nonetheless, the latter process could significantly affect the population ratios we observe today and obscure the optimistic constraints imposed by planetary migration models based on current resonant population numbers and distributions.

In a previous work, we showed that dwarf-planetary-sized perturbers could contribute to the resupplying of MMRs in cold debris disks, increasing the injection rate of low-inclination comets \citep{Munoz18}. In the solar system, the 34 largest TNOs can increase the number of Jupiter Family Comets (JFCs) injected into the inner Solar System \citep{Munoz19}. In that work, we did not observe significant variation in the evolution of the resonances, except for Pluto in the 3:2 MMR, where an increase in JFCs is evident \citep[see Figure 5 in][]{Munoz19}. This indicates that more objects are lost from the 3:2 MMR when Pluto is present compared to when it is absent. Given that the then-known 34 most massive objects do not directly influence any other resonances, we question whether less massive objects, but closer to the resonances, might have a more substantial impact on their evolution. This work aims to establish limits on this question using high-resolution numerical simulations.

In this work, we study the effect of the largest members inside Neptune's first-order MMRs in the KB, namely the 3:2 and the 2:1 MMRs, to determine their secular contribution to the evolution of the population size of such resonances after the end of the planetary migration phase.

This paper is organized as follows. Section \ref{sec:data} presents our two data sample sources. Section \ref{sec:simulations} describes our simulations and the initial characterization of potentially long-term resonant objects. Section \ref{sec:results} shows the behavior of the resonant populations, particularly their leaking rates as a function of time. Finally, our conclusions are presented in Section \ref{sec:conclusions}. 

\section{Data} \label{sec:data}

\begin{table*}[htbp!]
  \centering
  \caption{The ten largest prospective Plutinos and prospective Twotinos by absolute magnitude.}
  \begin{tabular}{lccccccl}
    \hline
    \hline
    Name & $H_V$ & $a$ (au) & p & $\rho$ (g/cm$^3$) & $R$(km) & $M$($\times10^{-4} M_{\earth}$) \\
    \hline
    \hline
    \multicolumn{7}{c}{Plutinos} \\
    \hline
    Pluto$^{\hyperlink{a32}{a}}$               & $-0.45$ & 39.49 & -  & 1.854 & 1188 & 24.467  \\
    Orcus$^{\hyperlink{b32}{b}}$               & $2.18$  & 39.28 & 0.23  & 1.676 & 450 & 1.073   \\
    Ixion$^{\hyperlink{c32}{c}}$               & $3.828$ & 39.50 & 0.14  & 1.28 & 309 & 0.263    \\
    \dpn{2003}{AZ}{84}$^{\hyperlink{d32}{d}}$  & $3.77$  & 39.47 & 0.11  & 0.87 & 386 & 0.349    \\    
    \dpn{2003}{VS}{2}$^{\hyperlink{e32}{e}}$   & $4.11$  & 39.42 & 0.15  & 1.19 & 318 & 0.266    \\
    \dpn{2003}{UZ}{413}$^{\hyperlink{f32}{f}}$ & $4.33$  & 39.23 & 0.14 & 1.05 & 241 & 0.104\\
    \dpn{2017}{OF}{69}$^{\hyperlink{f32}{f}}$  & $4.37$  & 39.44 & 0.14 & 1.03 & 237 & 0.097\\
    Huya$^{\hyperlink{f32}{f}}$                & $4.81$  & 39.41 & 0.14 & 0.86 & 193 & 0.044\\ % huya \dpn{2000}{EB}{173}
    \dpn{2002}{XV}{93}$^{\hyperlink{f32}{f}}$  & $4.88$  & 39.31 & 0.14 & 0.83 & 187 & 0.039\\
    Lempo$^{\hyperlink{f32}{f}}$               & $4.94$  & 39.41 & 0.14 & 0.81 & 182 & 0.039\\ % Lempo \dpn{1999}{TC}{36}
    \hline
    \hline
    \multicolumn{7}{c}{Twotinos} \\
    \hline
    \dpn{2002}{WC}{19}  $^{\hyperlink{f32}{f}}$  & $4.67$ & 47.86 &  0.14 & 0.91 & 207 & 0.056 \\
    \dpn{2005}{CA}{79}  $^{\hyperlink{f32}{f}}$  & $5.20$ & 47.78 &  0.14 & 0.73 & 162 & 0.021 \\
    \dpn{2012}{JH}{67}  $^{\hyperlink{f32}{f}}$  & $5.51$ & 47.78 &  0.14 & 0.63 & 140 & 0.012 \\
    \dpn{2021}{LN}{43}  $^{\hyperlink{f32}{f}}$  & $5.52$ & 47.73 &  0.14 & 0.63 & 140 & 0.012 \\
    \dpn{2007}{PS}{45}  $^{\hyperlink{f32}{f}}$  & $5.52$ & 47.56 &  0.14 & 0.63 & 140 & 0.012 \\
    \dpn{2015}{BE}{519} $^{\hyperlink{f32}{f}}$  & $5.60$ & 47.71 &  0.14 & 0.61 & 135 & 0.010 \\
    \dpn{2014}{DO}{143} $^{\hyperlink{f32}{f}}$  & $5.69$ & 47.58 &  0.14 & 0.58 & 129 & 0.008 \\
    \dpn{1998}{SM}{165} $^{\hyperlink{f32}{f}}$  & $5.78$ & 47.65 &  0.14 & 0.56 & 124 & 0.007 \\
    \dpn{2014}{WT}{69}  $^{\hyperlink{f32}{f}}$  & $5.89$ & 47.61 &  0.14 & 0.53 & 118 & 0.006 \\
    \dpn{2001}{UP}{18}  $^{\hyperlink{f32}{f}}$  & $5.98$ & 47.77 &  0.14 & 0.51 & 113 & 0.005 \\
    \hline
  \end{tabular}
  \label{tab:plutinos_twotinos}

  \hypertarget{a32}{a}: Mass of a single object at the barycenter, considering Charon's contribution. \cite{Stern18} \hfill \null\\
  \hypertarget{b32}{b}: Radius and density known from \cite{Barr16}. \hfill \null\\
  \hypertarget{c32}{c}: Only radius known from \cite{Lellouch13}; density and mass derived from \cite{Munoz19}. \hfill \null\\
  \hypertarget{d32}{d}: Radius and density derived from \cite{Mommert12,Dias17}. \hfill \null\\
  \hypertarget{e32}{e}: Only radius known from \cite{Mommert12}; density and mass derived from \cite{Munoz19}. \hfill \null\\
  \hypertarget{e32}{f}: All values derived using Eqs. 4 to 6 in \cite{Munoz19}. \hfill \null\\  
\end{table*}

\subsection{Initial Retrieval of Resonant Populations}

In this work, we explore the long-term evolution of the two strongest MMRs in the KB, namely the 3:2 and 2:1 first-order resonances with Neptune, which together account for approximately two-thirds of the observed population in all MMRs \citep{Gladman12,Forgacs23}. We used two different sources of data to define initial, potentially long-term, librating populations inside these resonances. The first set of data was obtained from observations and comes from NASA JPL's Small-Body Data-Base (SBDB)\footnote{\url{https://ssd.jpl.nasa.gov/tools/sbdb_query.html}}, as well as from The International Astronomical Union Minor Planet Center (MPC)\footnote{\url{https://www.minorplanetcenter.net/data}}. For this observational sample, we retrieved data from NASA JPL's SBDB, applying generous and straightforward criteria for the semimajor axis, $a$, selecting objects in the interval $38.8<a<40$ au for potential members of the 3:2 MMR and within $47<a<48.5$ au for potential members of the 2:1 MMR. These selection criteria yield the orbital parameters of 563 potential Plutinos (in the 3:2 MMR) and 153 potential Twotinos (in the 2:1 MMR).

The second source of population data is the theoretical model obtained from the Canada - France Ecliptic Plane Survey (CFEPS), the so-called L7 model\footnote{\url{https://www.cfeps.net/?page_id=105<}} \citep{Petit11,Gladman12}, which represents a debiased distribution of KBOs with magnitudes below $H_g\leq8.5$; the L7 model lists objects in different families in the trans-Neptunian region, namely several MMRs, the classical belt, and the scattered disk.

It is important to note that the L7 model is not intended to provide a comprehensive characterization of resonant populations. Instead, its goal is to reproduce the CFEPS observations after accounting for the well-understood survey biases, meaning it was designed to produce population estimates rather than serve as initial conditions for dynamical studies. However, given the L7 model's clear ability to reasonably match observations, we have decided to consider this population as a potential source of resonant particles. This choice is supported by the understanding that the model provides a fairly acceptable representation of resonant populations. In any case, after a thorough initial characterization, we will only be left with particles that exhibit libration over an initial simulation period of 10 Myr, a procedure similar to what we perform in our previous work on the delivery of Jupiter-family comets \citep{Munoz19}. 

From the L7 model, we obtained a large test particle sample containing the orbital parameters of 3340 and 871 objects inside the 3:2 and 2:1 MMRs, respectively. We did not use our semimajor axis criterion to filter L7 model data, as CFEPS already provides a classification of their particles; their filters include restrictions on the semimajor axis ranges of $39.25<a< 39.65$ au for the 3:2 MMR and $47.6<a< 48.0$ au for the 2:1 MMR.

Both of the above samples, SBDB and L7 populations, will be treated as test particles in our simulations.

\subsection{Orbital elements of the Giant Planets}

Our initial Solar System model consists of the Sun, to which we added the masses of the terrestrial planets and the Moon, as well as the four giant planets, treated as massive objects.

In this work, we performed a large number of numerical simulations with various initial test particle populations. Specifically, for the simulations using the L7 model to populate the resonances, we obtained precise heliocentric data for the giant planets (Jupiter, Saturn, Uranus, and Neptune) from JPL's NASA Horizons system
\footnote{\url{https://ssd.jpl.nasa.gov/horizons/app.html}}, for the Julian Date (JD) JD2453157.5, which corresponds to June 1, 2004, the epoch used for the L7 model generation, which lies at the middle of the CFEPS survey. Meanwhile, the SBDB objects are integrated starting at the epoch JD2460221, corresponding to October 3, 2023. For these simulations, we obtained precise heliocentric data for Jupiter, Saturn, Uranus, and Neptune from Horizons on the same date, JD2460221.

\subsection{The Sample of the Ten Most Massive Plutinos and Twotinos}
\label{sec:massive_DPs_sample}

Since a primary goal of this work is to characterize how large resonant bodies influence the secular evolution of the resonance as a whole, we select the ten most massive objects within the resonances to serve as massive perturbers in long-term simulations, alongside the giant planets.

From the observational sample, retrieved considering a large semimajor axis range, we identified the largest potentially resonant objects, which we will call prospective Plutinos and Twotinos, as those objects with the smallest absolute magnitudes. From these, we select the ten smallest $H_V$ objects that, in an initial 10 Myr integration under the perturbation of the giant planets and the Sun, show consistent libration in either resonance.

Our list of the 20 largest objects can be found in Table \ref{tab:plutinos_twotinos}. All these 20 identified objects coincided with the classification of \cite{Volk24}, i.e., they are classified as resonant based solely on 10 Myr integrations under perturbations from the Sun and the four GP. Among the set of the ten massive prospective Plutinos is, of course, Pluto, and other well-known Plutinos. Other objects in the prospective massive Plutino group, as well as all of the prospective massive Twotino group, are not so well characterized. 
This means that, from the 20 objects, only three of them have known masses: Pluto, Orcus, and \dpn{2003}{AZ}{84}. Of the other 17, two objects have known radius (Ixion and \dpn{2003}{VS}{2}), while for the other 15, only the absolute magnitude is known. To study the dynamical effect these objects would have on the MMR populations, we need to have at least an approximate determination of the mass. We do this using the fitting procedure of \citet{Munoz19}, where unknown albedos are assigned as a function of absolute magnitude, $H_V$, as follows:
\begin{equation}
p=
\begin{cases}
0.040H_V^2-0.259H_V+0.556 & \text{if } H_V \le 3.21\textrm{mag} \\
0.140 & \text{if } H_V > 3.21 \textrm{mag}.
\end{cases}
\label{eq:pvsh}
\end{equation}
Since all our objects without precise data have $H_V>3.21$, we assign a constant albedo of 0.140 to most of the objects in our list. We then obtain a radius following \citep{Harris97} as,
\begin{equation}
R=664.5 \textrm{km}\, {p}^{-0.5}10^{-H_V/5}.
\label{eq:dph}
\end{equation}
Finally, we assign a density as a function of radius following again the fitting from \citet{Munoz19} given by:
\begin{equation}
\rho=\left[\left(\frac{R}{220 {\rm km}}\right)^{-3}+(2.1)^{-3}\right]^{-1/3} \textrm{g/cm}^3,
\label{eq:rvsr}
\end{equation}
Equations \ref{eq:pvsh} and \ref{eq:rvsr} were determined by fitting the data of 27 objects with known albedo, and 13 objects with known density, respectively. For details about this procedure, see Appendix A in \citet{Munoz19}.

These objects will be used as a set of massive perturbers in later simulations to investigate how the largest resonant objects affect the leakage rates of the 3:2 and 2:1 MMRs.

\begin{figure}[htb!]
    \centering
    \includegraphics[width=\linewidth]{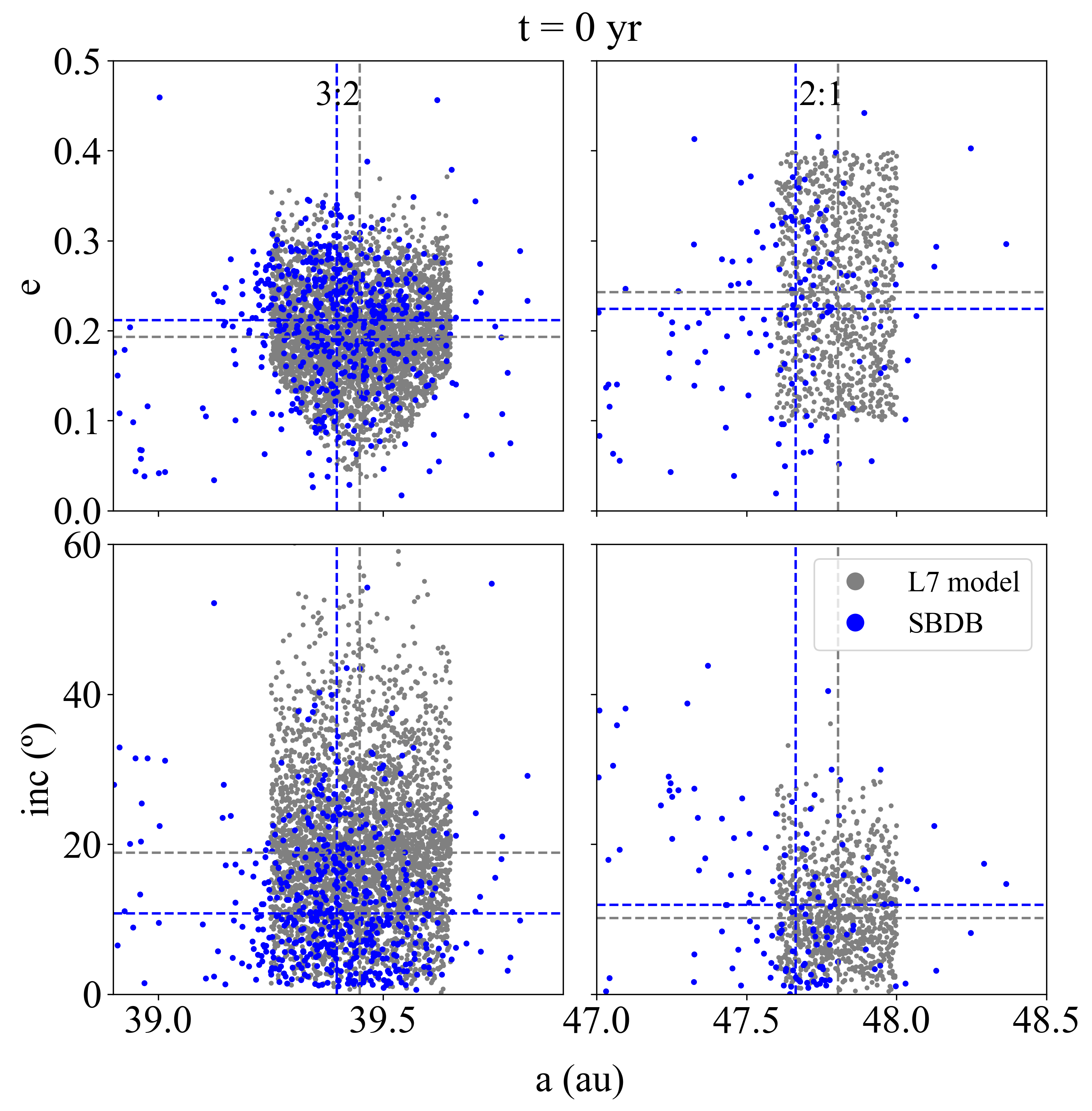}
    \caption{Initial orbital parameters of retrieved objects in the 3:2 (left panels) and 2:1 (right panels) MMRs. The top panels show the distribution in semimajor axis vs. eccentricity, while the lower panels show semimajor axis vs inclination. Gray dots show the sample obtained from the L7 model; blue dots show the observational sample obtained from JPL's SBDB. An initial statistical characterization is shown in dashed lines, which represent the median values of semimajor axis (vertical dashed lines) and eccentricity and inclination (horizontal dashed lines) for the L7 and SBDB populations, color-coded in the same way as the dots.}
    \label{fig:aei_0}
\end{figure}

\subsection{Initial Orbital Distribution of Resonant Populations}

Figure \ref{fig:aei_0} shows the initial orbital elements for the two test particle datasets (SBDB and L7) in the 3:2 and 2:1 MMRs. The left panels show the phase-space distribution ($a$ vs. $e$, top panel, and $a$ vs. $i$, bottom panel) of the 3:2 MMR, while the right panels show the distribution in the same planes of the 2:1 MMR populations. We show the initial distribution of the theoretically modeled L7 populations as gray dots, while blue dots represent the distribution of observed objects. We note that, for consistency, we integrate the L7 particles up to the same epoch as the SBDB sample, which is up to JD2460221, or roughly 19 years.

Some differences can be immediately highlighted between the two sets. On one hand, the eccentricity distribution of the L7 particles is initially symmetric in $a$ for both resonances, whereas observed particles tend to concentrate toward the lower semi-major axes. The medians of the semimajor axes of L7 particles (shown in vertical dashed gray lines) are 39.45 au and 47.80 au for the 3:2 and 2:1 MMRs, respectively; slightly larger than the medians of 39.40 au and 47.66 au found for the observed sample (shown in vertical dashed blue lines) which, in turn, are nearly identical to the average nominal MMRs, 39.40 au and 47.73 au, respectively, since $\bar{a}_{\it Nep}=30.07$ au is the average value of Neptune's semimajor axes in our simulations.

Eccentricities cover approximately the same range in both samples, having median values of 0.193 and 0.243 for the L7 model and 0.212 and 0.224 for the observational sample, for the 3:2 and 2:1 MMRs, respectively. 

Regarding inclination, the L7 model shows a considerably higher median for the 3:2 population compared to the SBDB, with values of 18.90$^\circ$ for the L7 model and 10.78$^\circ$ for the SBDB. This is likely due to the SBDB sample exhibiting strong observational biases that preferentially favor the discovery of objects with low inclinations. Conversely, the difference in medians for the 2:1 population is less pronounced, with values of 10.18$^\circ$ for the L7 population and 11.92$^\circ$ for the SBDB; this effect is due to an intrinsically tighter inclination distribution of the 2:1 MMR as well as an increased instability for higher inclination objects.

In the case of the 2:1 MMR, the spread of the observed sample in the semimajor axis is evident due to our ample selection criteria. We made this decision considering that the form of the resonances is not stringent (as suggested by the L7 model distribution) and does not follow simple analytical approximations, especially at low eccentricities and inclinations \citep[see, e.g.][]{Malhotra23}. In any case, to remain only with truly resonant particles, we will perform an initial characterization of the population samples shown in Fig. \ref{fig:aei_0}.

\section{Simulations: Initial filtering of resonant objects} \label{sec:simulations}

To filter out highly unstable objects in our sample and remain only with potentially long-term stable resonant objects in both populations (theoretical and observed) within the two resonances (3:2 and 2:1), short-term simulations were conducted with the numerical N-body integrator REBOUND \citep{rebound}. In all our simulations, we used MERCURIUS \citep{reboundmercurius} with a tolerance parameter of $10^{-9}$, a hybrid integrator that changes between a Wisdom-Holman symplectic mapping and the high-order integrator IAS15 \citep{Rein15} when solving close encounters with massive bodies. The initial characterizing simulations considered the Sun (with the mass of the terrestrial planets and the Moon added to it), the four giant planets, and the massless particles corresponding to our TNO samples. We used a timestep of 36.5 days and an output cadence of 500 yr, with $2\times10^{4}$, outputs, for a total integration time of 10 Myr.

\begin{figure}[htb!]
    \centering
    \includegraphics[width=\linewidth]{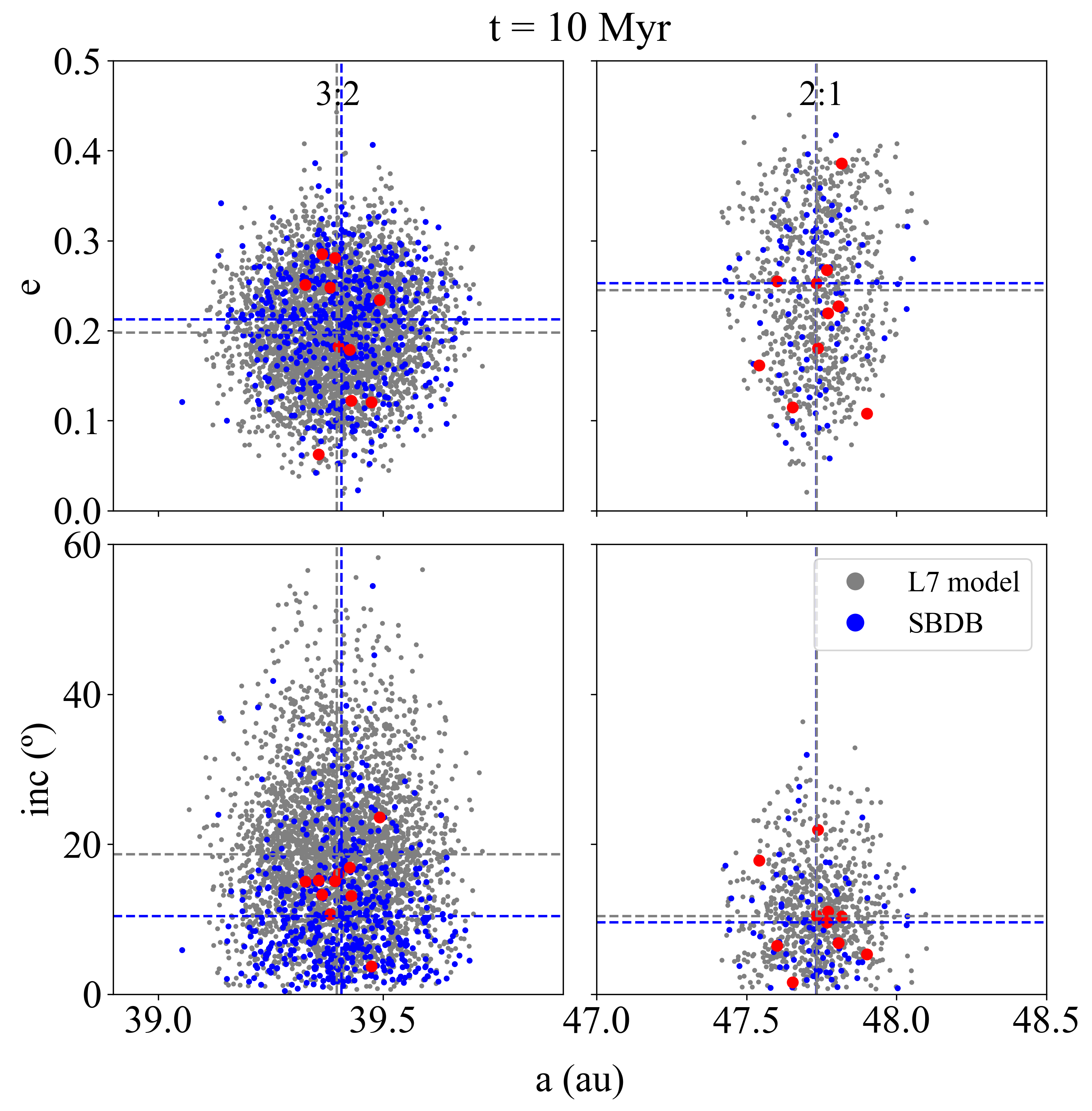}
    \caption{Orbital parameter distribution of resonant particles in our samples after a short-term, 10~Myr integration under perturbations from the Sun and the four giant planets only. As in Fig. \ref{fig:aei_0}, top panels show $a$ vs. $e$, and lower panels $a$ vs. $i$ distributions; also, left panels correspond to the 3:2 resonant populations and right panels to the 2:1 populations. As in Fig. \ref{fig:aei_0} L7 objects are plotted in gray, SBDB objects, however, are now divided into two groups where most of the objects are plotted in blue, but the ten most massive objects in each resonance are plotted in red; we will use these ten objects as massive perturbers in some of our long-term simulations, see text. Finally, dashed horizontal and vertical lines show median values of each element, color-coded as the dots, as in Fig. \ref{fig:aei_0}. All the panels show significant sculpting of the resonant populations, but this is especially evident in the $a$-$e$ plane.}
    \label{fig:aei_10}
\end{figure}

To identify potential stable members within the resonances, we analyzed for librating behavior the following resonant angles\footnote{We also analyzed other possible first-order arguments as given in \citet{MurrayDermott99}, involving the longitudes of pericenter, $\varpi$, and longitudes of the ascending nodes, $\Omega$, both for Neptune and the particle, corresponding to eccentricity and inclination type resonances, however, we did not find other librating arguments for either resonance.}:
\begin{equation}
\phi_{3:2} = 3\lambda_p - 2\lambda_N - \varpi_p,
\end{equation}
and 
\begin{equation}
\phi_{2:1} = 2\lambda_p - \lambda_N - \varpi_p,
\end{equation}
corresponding to the main resonant angles for the 3:2 and 2:1 MMRs, respectively. In the above equations, $\lambda_N$ corresponds to the mean longitude of Neptune, while $\lambda_p$ and $\varpi_p$ are the mean longitude and longitude of the perihelion of the particle, respectively. Objects with resonant angles librating with total amplitudes below 340$^\circ$ for the entire 10 Myr integration were classified as resonant. Under this criterion, the JPL's SBDB sample simulations yielded 500 and 97 resonant objects for the 3:2 and 2:1 populations, respectively. For the L7 model data, the numbers are 3161 and 771, respectively, for the 3:2 and 2:1 MMR. This is in excellent agreement with the characterization performed by \cite{Munoz19} on the same L7 data. The resulting orbital elements of these simulations are shown in Figure \ref{fig:aei_10}, with the same format used in Fig. \ref{fig:aei_0}. 

Fig. \ref{fig:aei_10} shows the effect of 10 Myr evolution on the particle distributions, where most of the long-term structure has begun to be imprinted. Dynamical sculpting in the current configuration of the solar system has recently been shown to shape some general properties of the observed distributions of resonant populations. However, the finer structure requires additional mechanisms to be reproduced appropriately \citep[see, for instance,][]{Balaji23}.  

In our short-term simulations corresponding to the 3:2 MMR (top-left panel in Fig. \ref{fig:aei_10}), both the L7 sample and the SBDB particles follow approximately the same semimajor axis vs eccentricity distribution after 10 Myr, having median $a$ values of 39.40 au and 39.41 au (nearly identical to the nominal value of the 3:2 MMR, as expected), and median $e$ values of 0.199 and 0.212 for the L7 and the SBDB samples, respectively. For inclination in the same 3:2 resonance (lower left panel), the L7 sample continues to show an overabundance of high-$i$ TNOs, which results from the same overabundance of high-$i$ objects in the initial distribution; median $i$ values after 10 Myr are found to be 18.58$^\circ$ and 10.43$^\circ$ for the L7 and SBDB samples, respectively. 

For the 2:1 MMR (right panels of Fig. \ref{fig:aei_10}), the small number of particles derived from observations does not allow us to make a significant comparison; however, the L7 sample seems to trace the expected shape for the resonant Twotino population adequately \citep[see e.g.][]{Robutel01,Munoz21}. On the other hand, most of the SBDB particles are found within this expected resonant region after 10 Myr. We found median semimajor axis values of 47.73 au and 47.73 au for the L7 and the SBDB samples (both identical to the nominal value of the 2:1 MMR). Median $e$ (top-right panel) and $i$ (bottom-right panel) values are found to be 0.245 and 0.253, and 10.24$^\circ$ and 9.61$^\circ$, again for the L7 particles and the observed SBDB particles, respectively. 

It is interesting to observe that the gravitational perturbations exerted by Neptune require less than 10 Myr to begin reshaping the original square distribution of the 2:1 population of the L7 model into an inverted teardrop with the vertex at $e=0$. The appearance of resonant particles with eccentricities below 0.1 in the evolution of the L7 2:1 sample does not diminish the median $e$ (in fact, it increases slightly), while the appearance of low $a$ objects does decrease the median $a$. The SBDB dataset, in contrast, increases in median $a$ due to the exclusion of a fraction of low $a$ objects outside the resonance. The net effect is that both samples now show identical $a$ medians at the location of the 2:1 MMR.

Fig. \ref{fig:aei_10} demonstrates that after just 10 Myr of evolution, the basic shape of the resonances has been dynamically sculpted by the giant planets. For our long-term simulations, we have highlighted some differences in the global properties of both populations, which deviate by small amounts; the most notable of these differences is an overabundance of high-$i$ particles in the 3:2 population of the L7 model, when compared to the SBDB, however, this discrepancy is most likely due to observational bias, where the L7 population try to correct for the presence of undiscovered objects at high-$i$. 

\section{Results and Discussion} \label{sec:results}

\subsection{Long-term simulations of the characterized resonant populations}
\label{subsec:long_term}

To explore the secular dynamics of the MMRs, particularly the leaking rate of particles eroding the resonant populations, we run long-term simulations using the subset of librating objects identified in our initial 10 Myr simulations. We ran these simulations from the end of the 10 Myr short-integrations to 4Gyr mark, using a 180 days and an output cadence of $2\times10^5$ yr, with 19,950, outputs, for a total integration time of $3.99\times10^9$ yr. We use these longer-spaced intervals to maintain a reasonable level of data output. Lastly, we ran an additional short-integration, identical to the one presented in Section \ref{sec:simulations}, from 4.0 Gyr to 4.010 Gyr to have high-frequency data at the end of the simulation for some of the distribution plots presented across Section \ref{sec:results} \citep[similar to the procedure of][]{Zhang22}. In all sets of simulations, the planetary energy is conserved at values below $10^{-6}$. The same MERCURIUS integrator from the REBOUND package was used for our long-term simulations with the same accuracy parameter as described in Section \ref{sec:simulations}.

\begin{figure}[htb!]
    \centering
    \includegraphics[width=\linewidth]{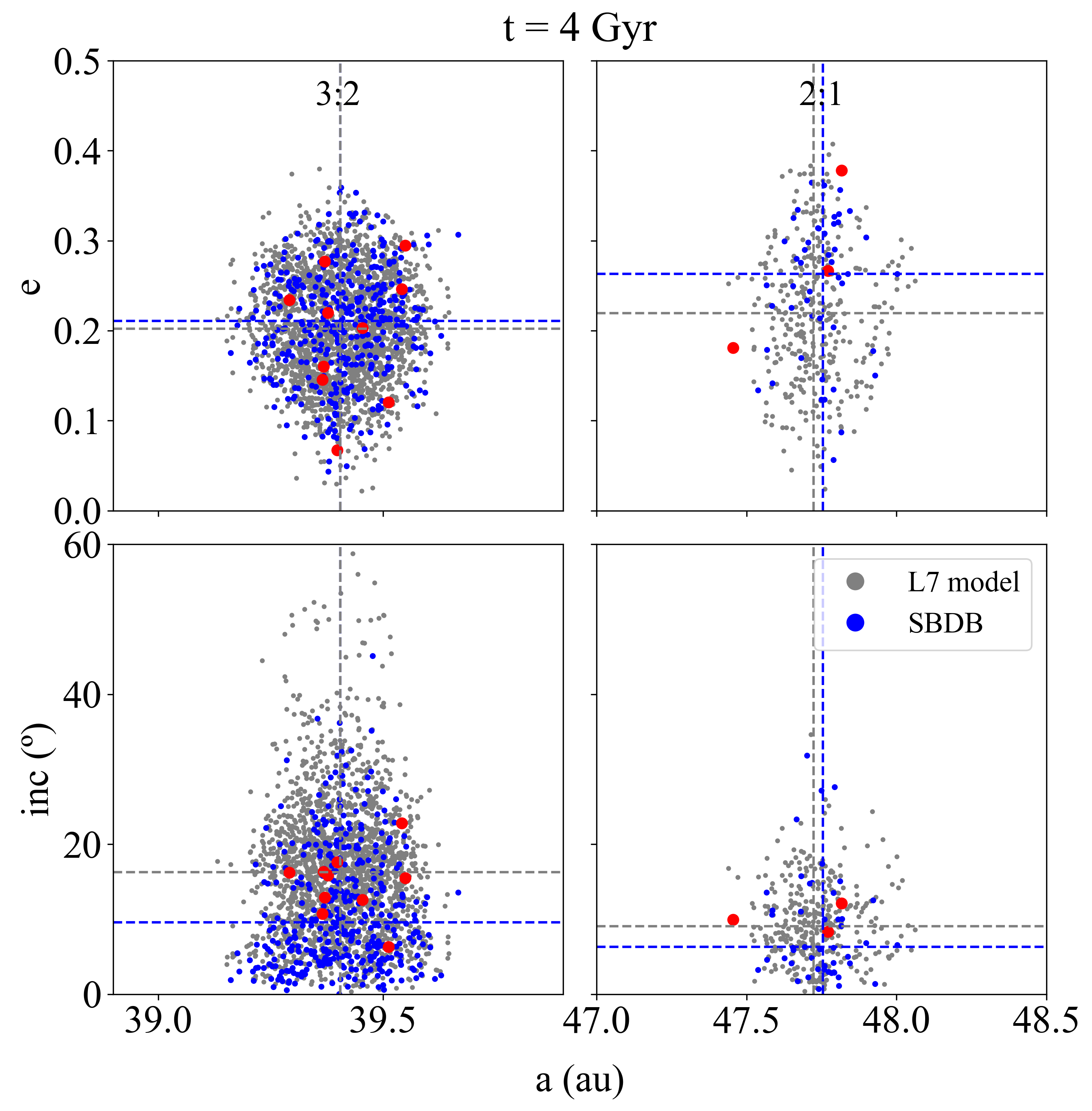}
    \caption{Same as Fig. \ref{fig:aei_10}, but for the orbital parameter distribution of resonant particles after a long-term, 4~Gyr integration under perturbations from the Sun and the four giant planets only.}
    \label{fig:aei_f}
\end{figure}

Figure \ref{fig:aei_f} follows the same format as Figs. \ref{fig:aei_0} and \ref{fig:aei_10}, but the distributions show particles that remain in resonance, i.e., they show consistent libration in the final 10 Myr short-integrations performed at the end of the long-term 4~Gyr simulation, with total amplitudes below 340$^\circ$. For the observed SBDB population (blue dots), our simulations produced 396 and 50 resonant objects inside the 3:2 and 2:1 MMRs, respectively. For the L7 model dataset (gray dots), 1919 and 330 objects remained inside the 3:2 and 2:1 MMR, respectively. The distributions of figure \ref{fig:aei_f} show solid sculpting of the 3:2 population \citep[in agreement with, e.g.][]{Balaji23}, where most of the scattered particles located away from the main bulk of the population have already been cleared out. A similar clearing occurs for the 2:1 populations, as we expected.

Although the medians fluctuate during the simulations, we use them to illustrate the differences between the populations we studied. At the end of the simulations, the median semi-major axis values for the remaining objects remain at 39.40 au and 47.75 au for the 3:2 and 2:1 SBDB populations, respectively. The L7 model shows a similar behavior, with values of 39.40 au and 47.72 au, which closely align with predictions based on Neptune's average semimajor axis. The median eccentricity values evolve to 0.211 and 0.263 for the 3:2 and 2:1 SBDB populations, respectively. For the L7 model, we obtain median eccentricity values of 0.202 and 0.219 for the 3:2 and the 2:1 MMRs. The inclination values do not converge as much as the other parameters. The median inclinations of the 3:2 and 2:1 observed populations are 9.60$^\circ$ and 6.31$^\circ$, respectively, while the L7 dataset final values are 15.65$^\circ$ and 9.07$^\circ$, respectively.

When studying the evolution of the Plutinos in the $a$ vs. $e$ plane (comparing the upper left panels of Figs. \ref{fig:aei_10} and \ref{fig:aei_f}), we see the occupied region has slimmed down, going from a bigger rounded shape, to a slightly slimmer "inverted-fat-teardrop" shape. This implies that objects in the outer part of the original shape are more likely to be lost from the resonance than those closer to the center, showing that objects with large eccentriciy ($e\gtrsim0.3$) or far away from the 3:2 MMR ($|a-39.40{\rm au}|\gtrsim0.2\,{\rm au}$) are less likely to remain in the resonance.

The $a$ vs. $e$  evolution of the Twotinos is less clear; overall, there is a clear loss of particles, but the overall size of the shape remains similar. 

%çççççççEste texto irá en lugar del parrafo anterior. .... The $a$ vs. $e$  evolution of the Twotinos is less clear; overall, there is a clear loss of particles \textbf{and, while the shape remains similar, it seems to slim down for $e>0.25$. The observed difference in the $a$ median values is not significant, as it is consistent with the statistical errors of the median determination; thus, it is not an indication of a difference between the two datasets. On the other hand, the decrease observed in the median value of $e$ in the L7 population indicates a significant feature, i.e., the sculpting of the L7 population remains ongoing even after 4 Gyr. For the SBDB, the small increase in median $e$ is not statistically significant and is consistent with a lack of evolution.}

Regarding inclination, we observe a consistent decrease in the median values for both the SBDB and L7 populations, as well as within both resonances. For the 3:2 MMR, the decline in the median value is less significant compared to that of the 2:1 MMR, where the inclination of the SBDB population is reduced to 66\% of its initial value. Specifically, the median value decreases from 9.61$^\circ$ at 10 Myr to 6.31$^\circ$ at 4 Gyr. In contrast, the L7 population shows a more pronounced decline in the 3:2 MMR than in the 2:1 MMR, with values decreasing from 18.58$^\circ$ at 10 Myr to 15.65$^\circ$ at 4 Gyr. These observations clearly indicate that particles with larger inclinations are more susceptible to perturbation and ejection from resonances, which aligns with previous findings \citep[e.g.][]{Nesvorny00,Tiscareno09}.

%*******çççççççç Según lo visto, las poblaciones de alta inclinación son considerablemente más inestables que aquellas de baja inclinación, en particular este efecto se presenta con una diferencia del 20% para particulas con inc>9.6 en SBDB. Debido a esto (aún considerando el rough initial sculpting del L7) es de esperar que la población debiased de L7 presente una estabilidad menor que la población del SBDB.

\begin{figure}[htb!]
    \centering
    \includegraphics[width=\linewidth]{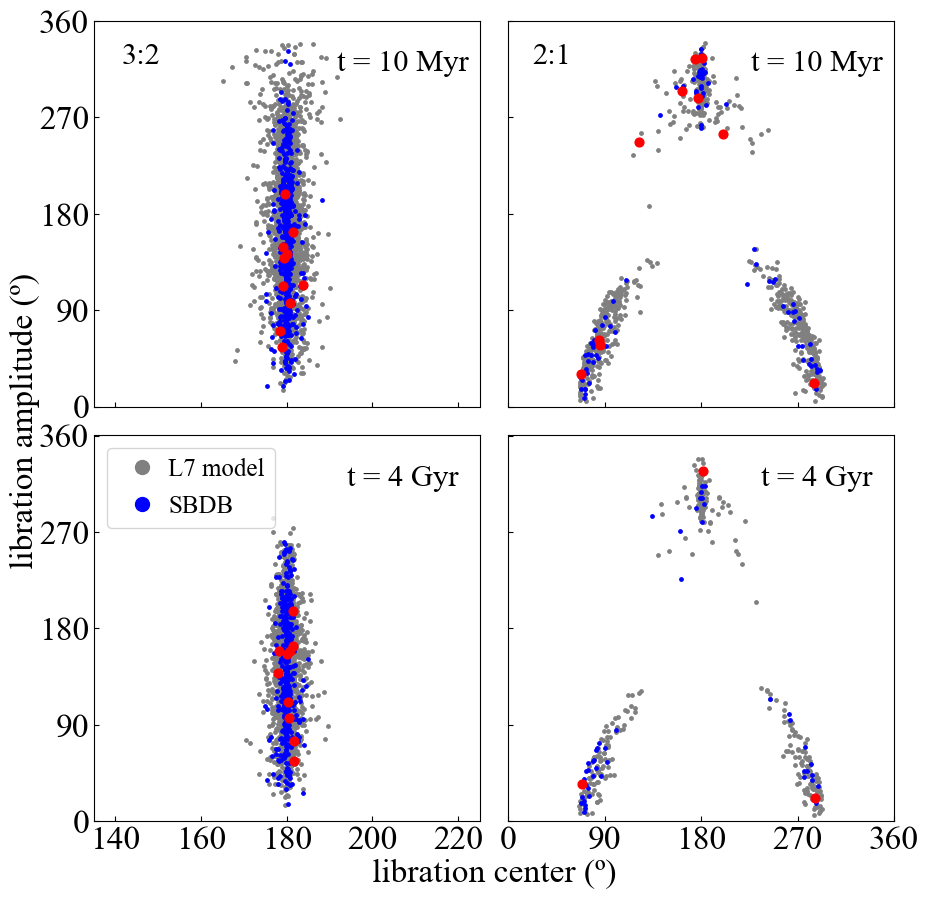}
    \caption{Initial (at 10 Myr, top panels) and final (at 4 Gyr, bottom panels) conditions for SBDB and L7 libration centers and libration amplitudes. The angles were calculated from the short-term integrations at the beginning and end of the 4 Gyr runs.}
    \label{fig:lib_center}
\end{figure}

To better visualize and compare the resonant characteristics of our samples and the effects of long-term evolution under the effect of the giant planets only, in Figure \ref{fig:lib_center}, we present the libration amplitude and the center of libration of the resonant angle of our studied populations; these angles were determined from the 10 Myr high-resolution simulations at the beginning (top panels) and end (bottom panels) of the 4 Gyr integrations, where libration centers are determined as the median values, and libration amplitudes are calculated using the maximum and minimum values of their oscillation.  

From Fig. \ref{fig:lib_center}, we see no major differences in the final distributions between the SBDB and L7 populations in both resonances. For both populations, the libration center of Plutinos (left panels) is symmetrically distributed around $180^\circ$, and there is no significant evolution, except for the removal of large libration amplitude members, those with total amplitude above $\sim260^\circ$. 

In the right panels of Fig. \ref{fig:lib_center}, the number of observed SBDB 2:1 MMR particles only allows for a partial comparison with the L7 population. Twotinos are distinguished by libration centers at $90^\circ$ (leading center), $180^\circ$ (symmetric center), and $270^\circ$ (trailing center). For the 2:1 MMR, again, no major changes are observed after 4 Gyr on the asymmetric islands, except for a reduction in the number of particles.

One striking feature in the evolution of the 2:1 population is the evolving relevance of the three libration centers. In the $\phi$ vs. $a$ plane, the leading and trailing centers appear as islands surrounded by the symmetric center, which in turn is bracketed by the area outside the resonance. Objects in the leading and the trailing centers are deep within the resonance, all having libration amplitudes less than $\sim180^\circ$, while those in the symmetric center are closer to the edges. As such, the symmetric center serves as the nexus, connecting to the other two libration centers and the surrounding population.  

For the SBDB, at the beginning of the simulation, the leading, trailing, and symmetric centers comprise approximately 35\%, 27\%, and 38\% (34, 26, and 37 particles) of the total population, respectively. By the end of the simulations, although all populations have decreased, the percentages have changed to 50\%, 26\%, and 24\% (25, 13, and 12 particles). The first thing to note is that the symmetric center has evaporated more rapidly than the other two. When examining the evolution of the asymmetric islands, we find that the trailing island appears to be evaporating slightly more rapidly than the leading one; however large, this difference is not statistically significant due to the small number of particles in our 2:1 population. While the most common explanation for the observed asymmetry between the leading and trailing islands is related to the capture efficiency within the 2:1 MMR during Neptune's early migration \citep[see e.g.,][]{Chiang02,Chen19}, this simulation suggests that the trailing island is more efficient at keeping objects from drifting into the symmetric island.

On the other hand, the L7 model does not delve deep into the nature of the particles trapped in the 2:1 resonance thus neither does it have some evolution history to shape its populations, nor does it look to reproduce the observed ratios between the leading and trailing islands; as such it starts with the leading, trailing, and symmetric centers containing 34\%, 36\%, and 30\% (262, 278, and 231 particles) of the total population, respectively; by the end of our simulations these fractions are 31\%, 30\%, and 39\% (104, 102, and 129 particles). The relevance of the symmetric center rises slightly, moving farther away from the observations. This occurs because at the beginning of the integrations, the leading and trailing centers are populated with many objects with relatively large libration (i.e., objects well bound to the resonance but weakly bound to their libration center, with relatively large libration amplitudes close to $140^\circ$). By the end of the simulation, fewer such objects remain; those that left the leading and trailing centers have migrated deep into the symmetric island, where they require more than 4 Gyr to escape efficiently from the 2:1 resonance. On the other hand, the ratio of leading to trailing populations is consistent with a constant 1:1 ratio, within errors, throughout the entire simulation; however, it is also consistent with a slightly faster evaporation of the trailing island, as was observed in the SBDB sample.

These results reinforce the caveat that caution should be exercised when applying the L7 model to study the evolution of specific aspects of the trans-Neptunian region \citep[as mentioned by the authors of the L7 model, e.g.][]{Gladman12}. Real objects tend to occupy the most stable orbits within a given area, whereas the fine details of L7 orbits are assigned randomly, placing them in relatively more "average" orbits. Additionally, there are many subdivisions within populations that have not been thoroughly considered, meaning that the population ratios will align with the total available phase space rather than with the most stable available phase space. Furthermore, while the long-term integration of sufficiently detailed models should help synthetic models to converge towards a more comprehensive representation of the Solar System, such convergence may not be straightforward and could go through spurious phases before achieving convergence.

The other noticeable feature lies in the fuzziness of the symmetric islands; while the leading and trailing centers appear to be sharp features, the symmetric center seems to have a halo underneath it; these points correspond to objects that are migrating between the symmetric island and one of the other two islands \citep[see also][]{Chen19}. The details of such transitions result in large libration amplitudes with intermediate libration centers.

\begin{figure}[htb!]
    \centering
    \includegraphics[width=\linewidth]{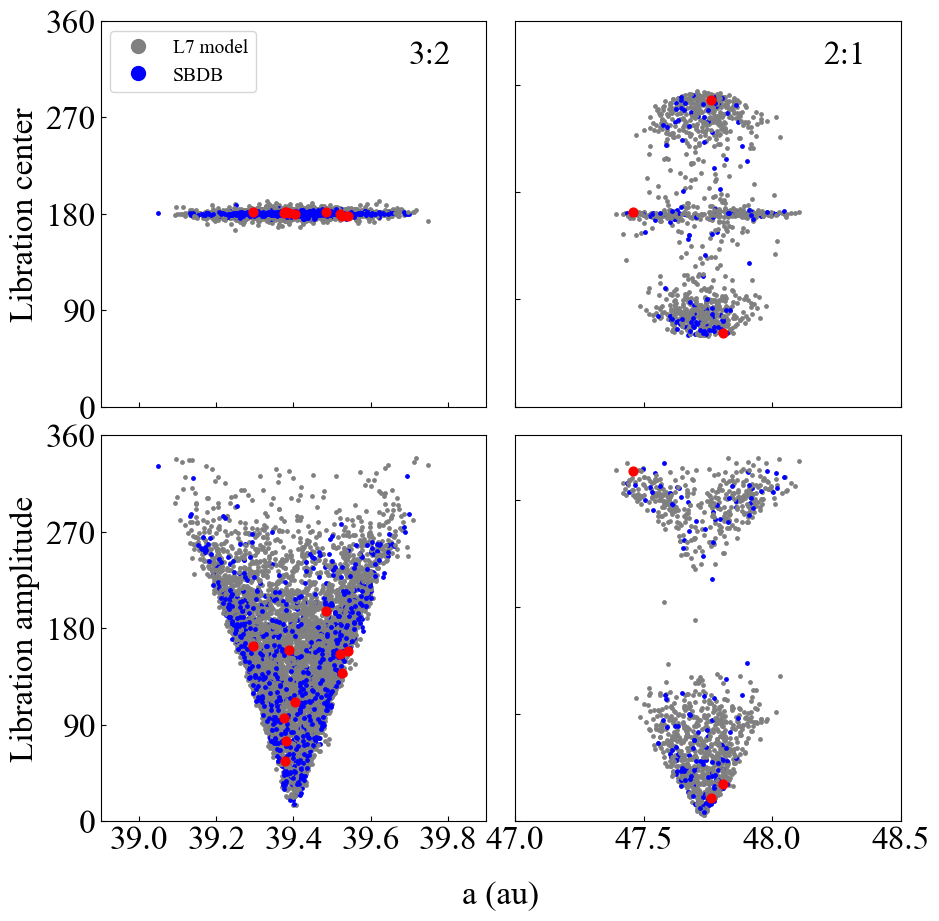}
    \caption{Distributions of libration centers and libration amplitudes of SBDB and L7 populations against their semimajor axes, at the end of long-term integrations. As in previous Figures, blue and gray dots indicate particles from the SBDB and the L7 models, respectively. The left column corresponds to the 3:2 resonant population, while the right column corresponds to the 2:1 population.}
    \label{fig:a_lib}
\end{figure}

Another way to look at the distribution of the 3:2 and 2:1 resonant families is shown in Fig. \ref{fig:a_lib}, where we present the amplitude and center of the resonant angles against the semimajor axis of each minor body at the 4 Gyr mark, again for the observed (blue dots) and L7 populations (gray dots). In the upper panels, $a$ vs. libration center, we observe a very compact distribution for the 3:2 population, whereas for the 2:1 population, three distinct concentrations are visible. These concentrations correspond to the leading, symmetric, and trailing distributions; since the concentrations overlap, this projection can not be used as a criterion to identify where each particle resides. On the lower panels, $a$ vs. libration amplitude, the 3:2 population shows that, for resonant particles, the lower the amplitude of its resonant argument, the closer the body lies to the center of the resonance at 39.40 au. A similar behavior is found for the 2:1 population, where particles closer to 47.73 au maintain lower libration amplitude values, although a considerable gap in the amplitude is found from 130$^\circ$ up to 225$^\circ$. As seen in figure \ref{fig:lib_center}, objects with libration amplitudes below this gap are part of the leading and trailing populations. In contrast, particles on the upper side of the gap correspond to those with a libration center at 180$^\circ$.

\begin{figure}[htb!]
    \centering
    \includegraphics[width=\linewidth]{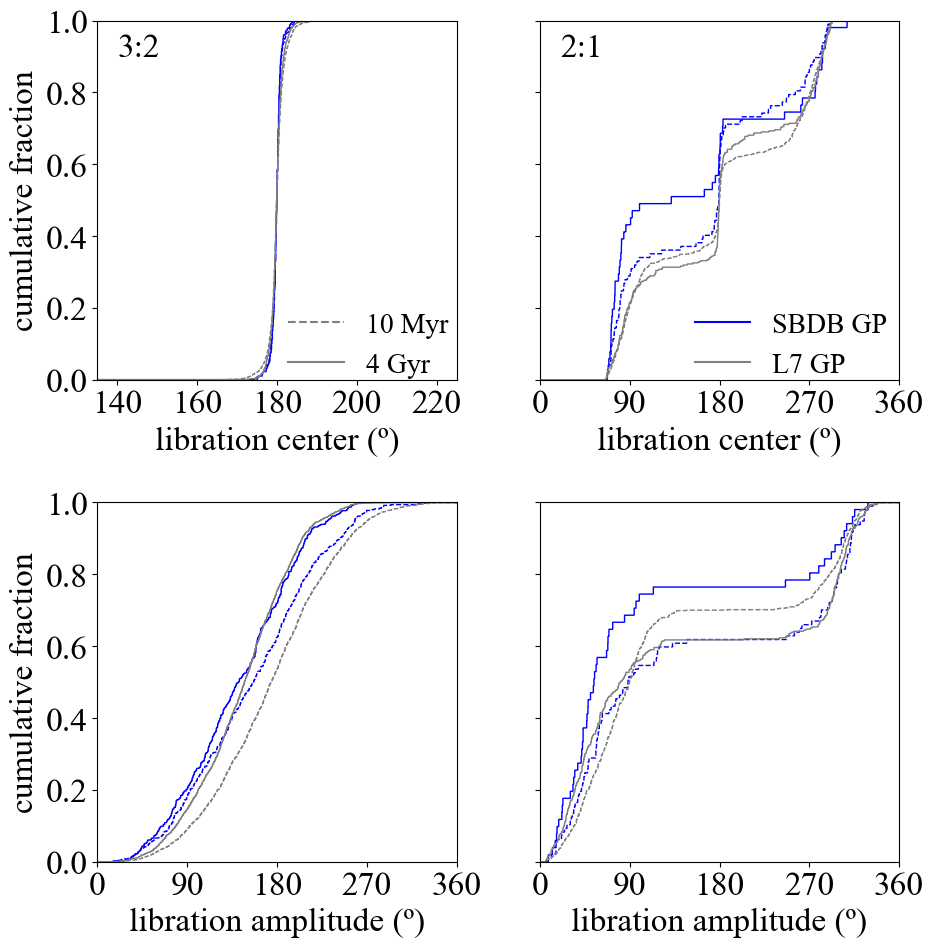}
    \caption{Cumulative fractions of the libration center (upper panels) and libration amplitude (lower panels) of the SBDB and L7 populations at the beginning and end of the simulations. Left panels show the cumulative fractions for the 3:2 MMR, while right panels show the cumulative fractions for the 2:1 MMR. Following the format of previous figures, the blue lines represent the SBDB, and the gray lines represent the L7 model. Dashed lines indicate the cumulative fractions at the beginning of the simulations, while solid lines denote such fractions after 4 Gyr.}
    \label{fig:cumm_fractions}
\end{figure}

The qualitative characteristics illustrated in Figs. \ref{fig:lib_center} and \ref{fig:a_lib} are furthermore analyzed in Fig. \ref{fig:cumm_fractions}, where we show the cumulative fraction of populations based on the libration center, in the upper panels, and the libration amplitude, in the lower panels. This representation is shown at the beginning and the end of the 4 Gyr simulations. As seen in previous figures, the data is divided between the 3:2 MMR (left panels) and the 2:1 MMR (right panels).

The libration center of the 3:2 resonance (upper left panel) exhibits a simple behavior that remains nearly unchanged, sharpening almost imperceptibly on Gyr timescales for both the SBDB and the L7 populations. The libration amplitude of the same 3:2 MMR (bottom left panel) is also very regular for both populations. Even though the initial distributions of both populations are quite different, after 4 Gyr the distributions are equivalent for particles with libration amplitudes above $\sim150^\circ$, reflecting the larger fraction of particles with lower libration amplitudes in the SBDB population.

Regarding the libration center of the 2:1 population (upper right panel), a clear difference is observed for the SBDB population after 4 Gyr. This is due to the substantial evolution of the leading island, which increases its fractional population by 15\%, while the symmetric island has the opposite effect. At the same time, the fractional population of the trailing island remains unchanged. On the other hand, since the biggest change in the fractional populations of the L7 model occurs in the symmetric island, it does not show as a significant jump in the evolution of the L7 model (gray) curves.  

% ççççççç Esta listo para integrarse al parrafo anterior.... \textbf{For the L7 model,} the biggest change in the fractional populations occurs in the symmetric island, \textbf{thus} it does not show as a significant jump in the evolution of the L7 model (gray) curves. \textbf{However, and despite 40\% of the twotinos being removed from the resonance, it is worth noticing that the 4~Gyr curve follows closely the 10~Myr curve up to $\sim100^\circ$ and again starting at $\sim260^\circ$ degrees up to $360^\circ$. This indicates that there is little evolution of the innermost parts of the asymmetric islands (those corresponding to particles with the smallest libration amplitudes, see Fig. \ref{fig:lib_center}) when comparing the initial and final states of the simulation. However, the outer parts of the asymmetric islands seem to be much less stable, with the corresponding particles now residing in the symmetric island. }

%  NOTA 1 la evolucion del SBDB sugiere que la ditribucion de las islas es un resultado de la evolucion que prefiere al leading; por otro lado el L7 sugiere que esto no es cierto y que probablemente venga de la formacion del SS. El texto que le mandamos al arbitro no refleja esto. Esto debe de dicutirse mas o menos aqui, y de reflejarse en las conclusiones.

Regarding the libration amplitude of the 2:1 MMR (lower right panel), we can see that most of the SBDB population in the asymmetric islands remains protected from ejections from the resonance, going from representing 60\% of the initial population to nearly 80\% by the end of the integrations. It is interesting to note that the trend of the L7 model moves in the opposite direction of the SBDB, with the population on the asymmetric island decreasing from 70\% at the start of the integration to approximately 60\% after 4 Gyr. Additionally, there is a notable similarity between the final L7 distribution of libration amplitude and the initial SBDB sample. 
%which suggests that the current shape of the solar system may influence the distribution of the libration island in the 2:1 MMR regardless of the initial state. 
As seen in Fig. \ref{fig:a_lib}, here we can again observe how there is a clear gap in libration amplitude between the symmetric and asymmetric islands; this is because the leading orbits have to limit themselves to instantaneous values of $\phi_{2:1}<180^\circ$ (most remain within $30^\circ<\phi_{2:1}<170^\circ$), trailing orbits have to limit themselves to $\phi_{2:1}>180^\circ$ (most remain within $190^\circ<\phi_{2:1}<330^\circ$); meanwhile, the symmetric island must encompass both other islands (whose exact location depends slightly on $e$ and $i$), resulting in a minimum libration in the $55^\circ<\phi_{2:1}<305^\circ$ range. The few objects in this gap correspond to objects that are switching between islands.

It is important to note that Twotinos with small libration amplitude, i.e., deeply trapped within the leading and trailing islands, remain trapped within their islands, librating in modes with low amplitudes. In contrast, objects located at the symmetric island may switch rapidly among the three modes.

\subsection{The evolution of the most massive Plutinos and Twotinos}
\label{subsec:massives}

Although Neptune’s gravity primarily shapes mean-motion resonances, the presence of other massive resonant bodies also influences the evolution of Plutinos and Twotinos. To quantify the effect that DPs have on resonant populations, we will first demonstrate how these objects influence the evolution of the ten most massive Plutinos and Twotinos when considered as massive perturbers in the simulations. This involves comparing the evolution of these objects when their self-interaction is included alongside perturbations from the giant planets.

The first step we need to take is to identify these 10 objects for long-term stability. Although we have shown that they remain within resonance for 10 Myr, the system is inherently chaotic, and there is no guarantee they will stay resonant throughout the entire 4 Gyr study; some dynamically perturbed DPs may be prone to drifting out of resonance. Furthermore, we test whether these ten objects remain in resonance with the presence of the additional perturbation provided by the mass of these same DPs.

Overall, we examined seven scenarios for the mass distribution in the MMRs, all including the perturbations of the sun and the four giant planets: (1) the 3:2 population as test particles; (2) the 3:2 population with the ten most massive Plutinos; (3) the 3:2 population perturbed only by Pluto; (4) the 2:1 population as test particles; (5) the 2:1 population with the ten most massive Twotinos; (6) the 2:1 population with the ten most massive Twotinos plus Pluto as an additional perturber; and (7) the 2:1 population perturbed only by Pluto, for comparison. 

Through our simulations, we identified which objects, characterized as the most massive resonant objects in section \ref{sec:massive_DPs_sample}, were long-term resonant, meaning their resonant argument remained librating with a full amplitude below 360$^\circ$ over a 4 Gyr integration period. Here we should note two differences from the full data-set analysis: 1) For the massive sample, we can examine each object visually; therefore, we can distinguish very high amplitude librations from circulations. This distinction is difficult to implement automatically, so we limit the libration angle criterion to a fixed 340$^\circ$ in the analysis of the full data set. 2) After the first 10 Myr, the cadence of our data points is 400 times slower, and some very fast events could be missed.

In Table \ref{tab:DPs_Classification}, we summarize our findings. Overall, we found that, under the perturbation of the GPs only, i.e., for the case of massless Plutinos, all of the sample remain resonant for the whole 4 Gyr integration, while when including the mass of the 10 most massive Plutinos (Pluto and the next nine most massive Plutinos) 9 of them were long-term resonant for the whole 4 Gyr, with one (\dpn{2003}{UZ}{413}) leaving the resonance at $\sim$1.5 Gyr; this object remained circulating close to the resonance for $\sim$100 Myr before being ejected from the solar system at $\sim$1.6 Gyr.

For the Twotinos, we found only two objects that are long-term resonant even in the massless DPs case, i.e., under the perturbations of the four GPs and the Sun. This highlights the importance of longer integrations for obtaining sounder classifications, for example, \dpn{2014}{DO}{143}, leaves the resonance very quickly, remains very close for 3.3 Gyr, and near the end of the simulation is recaptured into resonance; three more objects leave the resonance at $\sim$0.2, 2.8, and 2.8 Gyr, remaining as dropouts for the rest of the simulation; and the other four leave the resonance at $\sim$ 0.4, 0.6, 0.8, and 3.6 Gyr but these are shortly after ejected from the solar system. It is worth noting that additional observations would also be beneficial in constraining the orbital elements; this is especially true for the Twotinos, for which having accurate orbits may be critical due to the complex structure of their resonant phase-space.

Additional complications arise when considering mass in the sample. When considering the set of the ten largest Twotinos as perturbers in our simulations, we found the sample became overall more stable, with four additional objects (two of the dropouts and two of the ejected) remaining intermittently resonant for the entire integration. The remaining two previously ejected objects are ejected again, as is the last of the dropouts, which was not ejected previously; the net effect is that seven objects are in resonance instead of the three that were present at the end of the massless integration.

Finally, the addition of Pluto as a massive object in the integration, together with the ten massive Twotinos, destabilizes the set as compared to the simulation with massive Twotinos only, ejecting one more object after 4 Gyr. There is one additional object, \dpn{2005}{CA}{79}, that leaves the resonance after $3990$ Myr, and within 15 Myr (by $4005$ Myr, it is already on a hyperbolic orbit running away from the solar system). The effect is intermediate, with five objects in the resonance after 4 Gyr. 

\begin{table*}[htbp!]
  {\centering
  \caption{Plutinos and Twotinos found to be securely resonant, non-resonant, or that have been ejected in our simulations despite their initial resonant state.}
  \begin{tabular}{lclcl}

    \hline
    \hline
    \multicolumn{5}{c}{Plutinos} \\
    \hline
    \hline
     & \multicolumn{2}{c}{Massless DPs case} & \multicolumn{2}{c}{Massive DPs case}\\
    Name & R. Time & Final Status & R. Time & Final Status  \\
    \hline
    Pluto               & 4.0 Gyr & Resonant & 4.0 Gyr & Resonant \\
    Orcus               & 4.0 Gyr & Resonant & 4.0 Gyr & Resonant \\
    Ixion               & 4.0 Gyr & Resonant & 4.0 Gyr & Resonant \\
    \dpn{2003}{AZ}{84}  & 4.0 Gyr & Resonant & 4.0 Gyr & Resonant \\
    \dpn{2003}{VS}{2}   & 4.0 Gyr & Resonant & 4.0 Gyr & Resonant \\
    \dpn{2003}{UZ}{413} & 4.0 Gyr & Resonant & 1.5 Gyr & Ejected at $\sim1.6$ Gyr \\
    \dpn{2017}{OF}{69}  & 4.0 Gyr & Resonant & 4.0 Gyr & Resonant \\
    Huya                & 4.0 Gyr & Resonant & 4.0 Gyr & Resonant \\
    \dpn{2002}{XV}{93}  & 4.0 Gyr & Resonant & 4.0 Gyr & Resonant \\
    Lempo               & 4.0 Gyr & Resonant & 4.0 Gyr & Resonant \\
%    \hline
%    \hline
      \end{tabular}
      \\
\hspace{-1.7cm}
\begin{tabular}{lllllll}
    \hline
    \hline
    \multicolumn{7}{c}{Twotinos} \\
    \hline
    \hline
     & \multicolumn{2}{c}{Massless DPs case} & \multicolumn{2}{c}{Massive DPs case} & \multicolumn{2}{c}{Massive Twotinos + Pluto}\\
    Name & R. Time & Final Status & R. Time & Final Status & R. Time & Final Status \\
    \hline
    \dpn{2002}{WC}{19}  & 3.2 Gyr \ta & Ejected at $\sim3.6$ Gyr  & 4.0 Gyr \ta & Resonant                   & 4.0 Gyr \ta  & Resonant                   \\
    \dpn{2005}{CA}{79}  & 0.4 Gyr     & Ejected at $\sim0.7$ Gyr  & 2.3 Gyr \ta & Ejected at $\sim2.5$ Gyr   & 3.99 Gyr \tc & Ejected at $\sim 4.01$ Gyr \tc\\
    \dpn{2012}{JH}{67}  & 4.0 Gyr     & Resonant                  & 4.0 Gyr     & Resonant                   & 4.0 Gyr      & Resonant                   \\
    \dpn{2021}{LN}{43}  & 2.8 Gyr     & Dropout, $a=47.7$         & 4.0 Gyr \ta & Resonant                   & 4.0 Gyr      & Resonant                   \\
    \dpn{2007}{PS}{45}  & 0.2 Gyr     & Dropout, $a=48.2$         & 0.6 Gyr \ta & Ejected at $\sim 1.3$ Gyr  & 0.9 Gyr      & Ejected at $\sim 1$ Gyr    \\
    \dpn{2015}{BE}{519} & 0.8 Gyr     & Ejected at $\sim0.8$ Gyr  & 0.6 Gyr     & Ejected at $\sim1.2$ Gyr   & 0.9 Gyr \ta  & Ejected at $\sim 1.2$ Gyr  \\
    \dpn{2014}{DO}{143} & 20  Myr \tb & Resonant -- Recaptured    & 60  Myr \tb & Resonant -- Recaptured     & 6 Myr   \tb  & Resonant -- Recaptured      \\
    \dpn{1998}{SM}{165} & 4.0 Gyr     & Resonant                  & 4.0 Gyr     & Resonant                   & 0.1 Gyr      & Ejected at $\sim 0.12$ Gyr \\
    \dpn{2014}{WT}{69}  & 0.6 Gyr     & Ejected at $\sim0.8$ Gyr  & 4.0 Gyr \ta & Resonant                   & 2.2 Gyr \ta  & Ejected at $\sim 3.9$ Gyr  \\
    \dpn{2001}{UP}{18}  & 2.8 Gyr \ta & Dropout, $a=47.6$         & 4.0 Gyr \ta & Resonant                   & 4.0 Gyr      & Resonant                   \\
    \hline
      \label{tab:DPs_Classification}
  \end{tabular}}
  \tablenotetext{a}{This object overall librates, but intermittently circulates during its residence time.}
  \tablenotetext{b}{Despite leaving the 2:1 MMR very quickly for the first time, and having complex intermittent periods of librations and circulations, in all three simulations \dpn{2014}{DO}{143} finishes with a long period in resonance (between 400-700 Myr, depending on the simulation).}
  \tablenotetext{c}{This object remained in the resonance for $3\,990$ Myr, and was soon placed in a hyperbolic orbit; this occurred at $4\,005$ Myr, just after the nominal end of our simulations.}
\end{table*}

%    \dpn{2002}{WC}{19}  & 3.2 Gyr *           & Ejected at $\sim3.6$ Gyr & Time & Fill status  & 4.0 Gyr *            & Resonant                    \\
%    \dpn{2005}{CA}{79}  & 0.4 Gyr \phantom{*} & Ejected at $\sim0.7$ Gyr & Time & Fill status  & 2.3 Gyr *            & Ejected at $\sim2.5$ Gyr    \\
%    \dpn{2012}{JH}{67}  & 4.0 Gyr \phantom{*} & Resonant                 & Time & Fill status  & 4.0 Gyr  \phantom{*} & Resonant   \\
%    \dpn{2021}{LN}{43}  & 2.8 Gyr \phantom{*} & Dropout, $a=47.7$        & Time & Fill status  & 4.0 Gyr *            & Resonant                    \\
%    \dpn{2007}{PS}{45}  & 0.2 Gyr \phantom{*} & Dropout, $a=48.2$        & Time & Fill status  & 0.6 Gyr *            & Ejected at $\sim 1.3$ Gyr   \\
%    \dpn{2015}{BE}{519} & 0.8 Gyr \phantom{*} & Ejected at $\sim0.8$ Gyr & Time & Fill status  & 0.6 Gyr  \phantom{*} & Ejected at $\sim1.2$ Gyr    \\
%    \dpn{2014}{DO}{143} & 20  Myr \phantom{*} & Resonant (Recaptured)    & Time & Fill status  & 60  Myr **           & Resonant (Recaptured) \\
%    \dpn{1998}{SM}{165} & 4.0 Gyr \phantom{*} & Resonant                 & Time & Fill status  & 4.0 Gyr  \phantom{*} & Resonant                    \\
%    \dpn{2014}{WT}{69}  & 0.6 Gyr \phantom{*} & Ejected at $\sim0.8$ Gyr & Time & Fill status  & 4.0 Gyr *            & Resonant                    \\
%    \dpn{2001}{UP}{18}  & 2.8 Gyr *           & Dropout, $a=47.6$        & Time & Fill status  & 4.0 Gyr *            & Resonant                    \\
    
% *** Leaves resonance at 3.98 Gyr and is ejected at 4.006 Gyr

\subsection{Leaking Rates with and without Massive Plutinos and Twotinos}

Resonances do not remain with constant populations throughout their history. Instead, a continuous leaking process keeps these populations in constant evolution, helping, for example, to feed the Jupiter Family comet population on secular time scales \citep[e.g.][]{Ip97,Morbidelli97,Munoz19}. Here, we analyze and compare the leaking rates of the resonances in the different scenarios considered. Simulations that include only the giant planets as gravitational perturbers serve as a point of comparison to simulations that include the ten most massive Plutinos and Twotinos, in addition to which we analyze the individual effect of Pluto on the evolution of both resonances, as well as together with the ten most massive Twotinos for the 2:1 MMR.

It is worth noticing that the acceptance and rejection criteria in long-term integrations are less precise than in short-term ones. To confidently determine that a circulating orbit will have a data point outside the 340$^\circ$ resonating window, more than 100 data points are needed. Given our output cadence, this means over 20 Myr. During these time scales, most orbits---even those in resonance---undergo some evolution, leading to false negatives. In contrast, short-term integrations do not face this issue since 100 data points cover only $50\,000$ years, which is quicker than typical orbital evolution. While long-term integrations can not be directly compared to short-term ones, the measurements remain overall consistent. The evolution of the resonant populations in long-term simulations thus allows us to accurately assess the fractional change over the 4 Gyr study period.

\begin{figure}[hbt!]
    \centering
    \includegraphics[width=\linewidth]{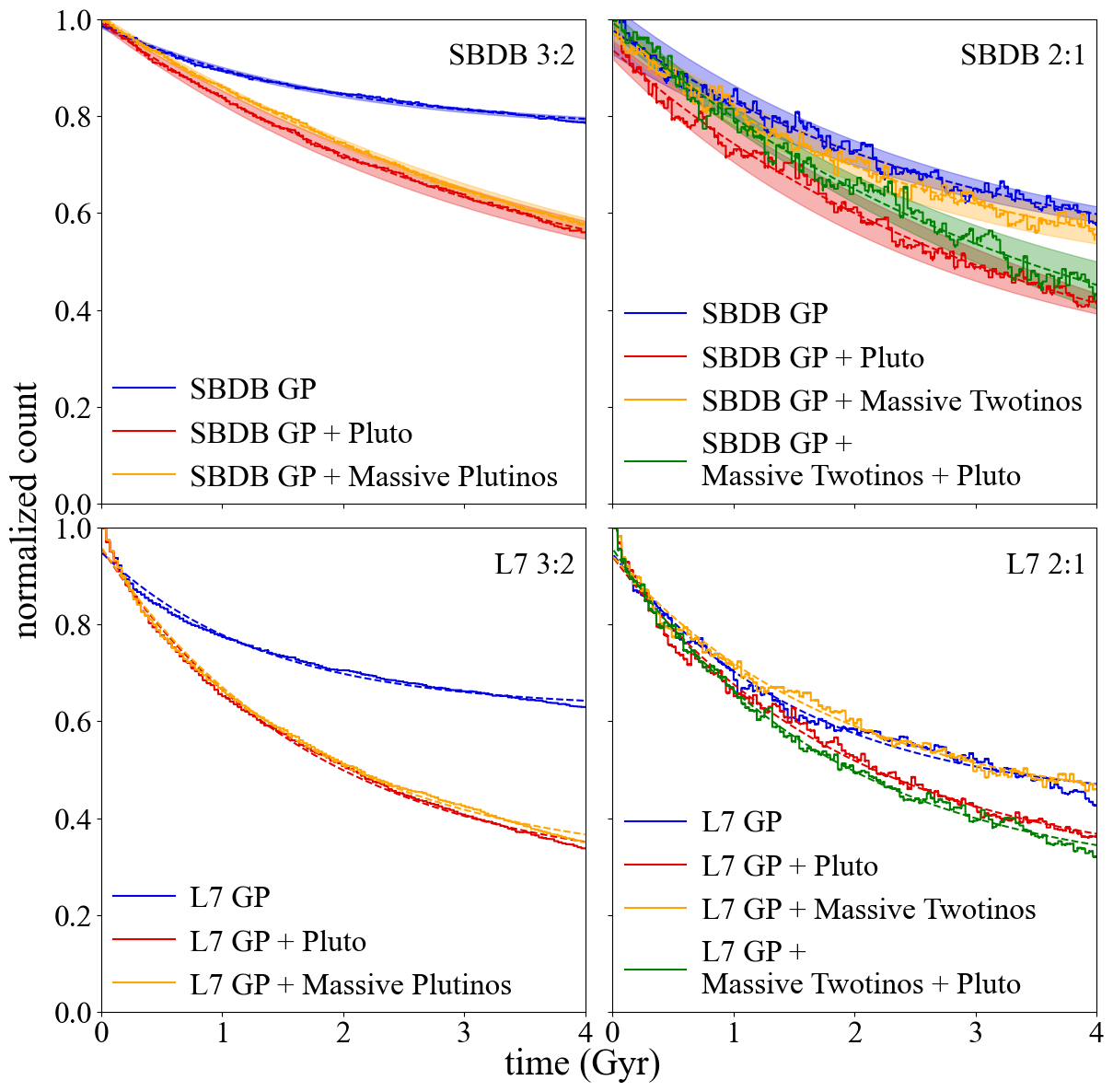}
    \caption{Evolution of normalized counts of resonant objects from SBDB (upper panels) and L7 (lower panels) samples. The left panels correspond to simulations of the 3:2 population, and the right panels to the 2:1 population. The solid color lines indicate the number of objects, while the dotted lines correspond to exponential fits as described in equation \ref{eq:leak_rate} and table \ref{main_fits} in Section \ref{sec:comparison_rates}. Normalization is done with respect to the number of objects at 10 Myr. The colored shadows in the upper panels represent the variations obtained from the 3 different runs we performed in each scenario for the SBDB sample. Lines corresponding to the L7 model (bottom panels) do not have shadows, as their simulations were conducted only once due to computational expense.}
    \label{fig:leak_rate}
\end{figure}

The fractional evolution of each population for the different cases considered is shown in Figure \ref{fig:leak_rate}. The two left panels of Fig. \ref{fig:leak_rate} correspond to the six scenarios studied for the 3:2 population, while the two right panels show the same six cases plus two scenarios of Pluto and the massive Twotinos for 2:1 populations. Furthermore, the upper two panels represent the SBDB dataset, while the bottom two represent the objects from the L7 model. Although the SBDB sample is relatively small, \citet{Munoz21} shows that, for orbits in chaotic phase space, each object evolves as a cloud of possibilities. By integrating the same orbit multiple times with different time steps, each particle unfolds into several possible evolutions. We use these results to enhance the statistical significance of the SBDB sample's outcomes by running them three times (with time steps of 180, 200, and 220 days). The larger L7 samples make multiple runs more costly, while providing only minor insights. The resulting variations in the leaking rates due to the time-step variations are negligible for the 3:2 population, as indicated by the shaded areas surrounding the blue, red, and orange curves in the upper left panel. In contrast, for the 2:1 population, these errors appear to be more significant, causing some overlapping of the shaded areas (color-coded in the same way as for the 3:2 case) corresponding to the evolution of the SBDB population with and without massive Twotinos. We also present the cases, including Pluto alone with the 2:1 population, as well as Pluto together with the ten massive Twotinos (green curve, a case only applicable to the 2:1 populations). The lower panels show the same cases, color-coded in the same way as the upper panels, but for the L7 populations.

The 3:2 population is greatly affected by the presence of Pluto, as evidenced by the significant gap between the blue and red curves in the upper left panel of Fig. \ref{fig:leak_rate}, which represents the evolution of the observed Plutino population with the giant planets only (blue) and the giant planets plus Pluto (red). On the other hand, the other nine massive Plutinos have a minimal effect on overall stability, as indicated by the orange curve. This effect is further confirmed by comparing the equivalent curves in the bottom-left panel, which corresponds to the evolution of the L7 Plutino population. 

In the case of Twotinos (right panels of Fig. \ref{fig:leak_rate}), the evolutionary track of simulations that included only the giant planets is very similar for both the SBDB and L7 populations (blue curves in the upper and lower right panels). Adding the ten massive Twotinos into the simulations (orange curves) results in only a slight difference compared to the giant planet case alone, i.e., the perturbations produced by the most massive Twotinos are almost negligible, as were the perturbations of the nine massive Plutinos besides Pluto for the 3:2 population. Surprisingly, a more critical instability on the 2:1 population is caused by Pluto alone, as its presence has a bigger effect than the ten massive Twotinos. This can be observed by comparing the red and green curves with the orange curve, and this effect is also evident in the L7 model, as shown by the corresponding curves in the bottom-right panel.

Pluto's influence on the 2:1 populations may be attributed to the dwarf planet's close encounters at aphelion with Twotino bodies. However, this suggests that Eris might also play an important role for Plutino bodies, given its perihelion is approximately 38 au. To explore this idea, we conducted numerical simulations using the 3:2 populations from the SBDB and L7 samples under perturbations from the giant planets and Eris. Our findings indicate that Eris has a minimal impact on the leak rate of the resonance, accounting for approximately 1\% more leaking, when compared to the case of the giant planets alone. Another possibility for Pluto's effect on the 2:1 population is their mutual 4:3 MMR. The significance of this resonance has not been previously explored, since Pluto is not massive enough to trap objects in its resonances. However, Neptune can trap Twotinos in Pluto's 4:3 resonance, where Pluto can perturb them secularly. We are exploring this idea in Ram\'irez-Vargas et al. in prep.

\subsection{Leaking rate fittings and comparison with previous works}
\label{sec:comparison_rates}

\begin{table*}[hbt!]
    \centering
    \begin{tabular}{lccc}
    \hline\hline
        Simulation & $\tau$ [Gyr] & $N_U^0/N_T^0$ & $N_S/N_T^0$ \\
        \hline
        \hline
        \multicolumn{4}{c}{Plutinos} \\
        \hline

        SBDB GP                    &  $1.90 \pm 0.03$ &  $0.22 \pm 0.01$ &  $0.77 \pm 0.01$ \\
        SBDB GP + Pluto            &  $3.58 \pm 0.34$ &  $0.64 \pm 0.01$ &  $0.36 \pm 0.01$ \\
        SBDB GP + Massive Plutinos &  $4.81 \pm 0.52$ &  $0.76 \pm 0.07$ &  $0.25 \pm 0.07$ \\
        L7 GP                      &  $1.33$ &  $0.32$ &  $0.63$ \\
        L7 GP + Pluto              &  $1.77$ &  $0.67$ &  $0.28$  \\
        L7 GP + Massive Plutinos   &  $1.61$ &  $0.64$ &  $0.32$  \\
        L7 GP + Massive Plutinos   &  $1.732 \pm 0.000$ &  $0.660 \pm 0.000$ &  $0.301 \pm 0.000$  \\
        \hline
        \hline
        \multicolumn{4}{c}{Twotinos}\\
        \hline
        SBDB GP                               &  $2.89 \pm 0.10$ &  $0.51 \pm 0.04$ &  $0.47 \pm 0.01$ \\
        SBDB GP + Pluto                       &  $3.35 \pm 1.03$ &  $0.75 \pm 0.12$ &  $0.19 \pm 0.10$ \\
        SBDB GP + Massive Twotinos            &  $2.67 \pm 0.25$ &  $0.53 \pm 0.01$ &  $0.45 \pm 0.01$ \\
        SBDB GP + Massive Twotinos + Pluto    &  $3.60 \pm 0.54$ &  $0.81 \pm 0.11$ &  $0.19 \pm 0.13$ \\
        L7 GP                                 &  $1.57$ &  $0.52$ &  $0.43$  \\
        L7 GP + Pluto                         &  $1.95$ &  $0.66$ &  $0.28$ \\
        L7 GP + Massive Twotinos              &  $1.59$ &  $0.51$ &  $0.43$ \\
        L7 GP + Massive Twotinos + Pluto      &  $1.80$ &  $0.69$ &  $0.27$ \\
        \hline
    \end{tabular}

    \caption{Parameters for the exponential fittings based on equation \ref{eq:leak_rate} for each set of simulations presented on Figure \ref{fig:leak_rate}. Note that the best numerical fit sometimes gives $N_U^0/N_T^0+N_S/N_T^0 \neq 1.0$.}
    \label{main_fits}

\end{table*}

In all the cases shown in Fig. \ref{fig:leak_rate}, the evolutionary tracks display a negative gradient that gradually decreases over the course of the simulation. We find that each track can be well described by a negative exponential with a constant offset, characterized by the fitting relations given in the following equations: 
\begin{equation}
    N_T(t) = N_U(t)+N_S = N_U^0\exp{\left(-\frac{t}{\tau}\right)} + N_S,
    \label{eq:leak_rate}
\end{equation}
and
\begin{equation}
    \dot{N_T}(t)=-\frac{N_U}{\tau},
\end{equation}
where $N_T(t)$ is the resonant population at time $t$, which can be divided into two populations, $N_S$ and $N_U$ representing the stable and unstable populations, respectively, with $N_U^0$ being the unstable population at $t=0$. Note that $N_S$ is a constant offset related to the stable fraction of particles in the resonance, and $\tau$ gives the exponential decay rate of the unstable population. This behavior is consistent with the stochastic nature of the leaking phenomenon from resonances, i.e., the escape of objects from the resonances appears to be driven by random processes, as indicated by this distribution. At the same time, the constant offset can be attributed to hyper-stable objects, such as Pluto \citep[e.g.][]{Ito25}, but it consists of a significant fraction of the original population. 

The characteristic time, or leakage rate, at which unstable objects escape from the resonances differs for both MMRs and both datasets, as shown in Table \ref{main_fits}, where we present the specific parameters of the exponential fittings to each simulation shown in Figure \ref{fig:leak_rate}. Note that we allowed the best numerical fitting, in which $N_U^0 + N_S$ may not add exactly to $N_T^0$. From Table \ref{main_fits} we see slower escape rates for the SBDB than for the L7 models, as well as slower rates for the 3:2 MMR than for the 2:1 MMR. For simulations with the giant planets only, the unstable population in the SBDB 3:2 MMR halves every 1.32 Gyr; meanwhile, the 2:1 unstable population halves every 2 Gyr.  

Another factor in the leakage is the ratio of the stable to unstable particles. Figure \ref{fig:leak_rate} and Table \ref{main_fits} show that all simulations without Pluto are more stable than the ones including Pluto, regardless of the nine additional massive Plutinos or the ten massive Twotinos. While this result may be expected for the 3:2 population, the effect of Pluto in the Twotino is quite unexpected. Quantitatively, in the absence of Pluto, the fraction of unstable particles lies in the 22\% to 53\% range, while including Pluto, the fraction of unstable particles rises to the range between 64\% to 81\%.

Other than that, the 3:2 MMR is generally more stable than the 2:1 resonance, with a fractional difference between unstable populations in the range of $0.15 \pm 0.16$. Particles from the SBDB model exhibit a larger stable fraction in the absence of Pluto, but a smaller fraction once Pluto is included. Nevertheless, the SBDB model shows slower exponential decay. Overall, across the 4~Gyr explored here, simulations with SBDB particles remain at least as stable---and in some cases up to 20\% more stable---than the corresponding simulations with L7 model particles. 

Though not many, there have been some previous quantifications of the leaking rates from Neptune's 3:2 and 2:1 MMRs. \citet{Morbidelli97} estimates a power law function of the time with an exponent of -0.5 for the 3:2 MMR, while \citet{Tiscareno09} found exponents of -0.55 and -0.77 for the 3:2 and 2:1 MMRs, respectively. Based on these works, \citet{Greenstreet15} estimated the impact rates on Pluto and Charon during the last 4 Gyr, using a power law fitting given by:
\begin{equation}
\frac{N(t)}{N_0}=\left(\frac{4.5\,\textrm{Gyr}}{t}\right)^b,
\label{eq:green}
\end{equation}
with $b=0.52$ for the 3:2 MMR and $b=0.77$ for the 2:1 MMR. \citet{Greenstreet15} defined equation \ref{eq:green} so that the present time corresponds to 4.5 Gyr; therefore,  to use it, we have to extrapolate it from 4.5 to 8.5 Gyr (i.e., we use it from the present time to 4 Gyr into the future). 

\begin{figure}[hbt!]
    \centering
    \includegraphics[width=\linewidth]{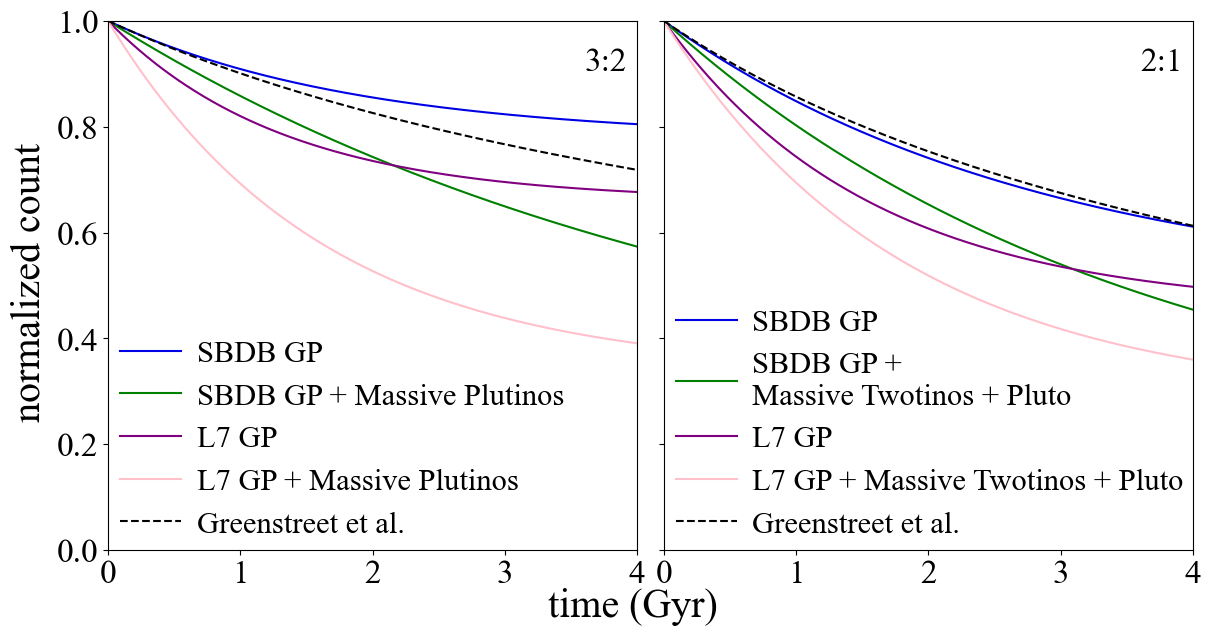}
    \caption{Leaking rate fit comparisons. Our rates (solid colored lines) come from Table \ref{main_fits}, but have been renormalized so they start with a value of 1.00 at $t=0$ Gyr. The dashed line corresponds to Eq. \ref{eq:green}, taken from \citet{Greenstreet15}.}
    \label{fig:comparison_leakings}
\end{figure}

In Fig. \ref{fig:comparison_leakings} we compare Eq. \ref{eq:green} with our exponential fittings for each resonance and for both the SBDB and L7 populations. Although \citet{Greenstreet15} models only consider the four giant planets as perturbers, in this comparison, we include simulations with only the giant planets as perturbers, as well as our most complete simulations; that is, in the case of Plutinos, the case including the ten massive Plutinos, and for Twotinos, the case including the ten massive Twotinos and Pluto. 

Overall, we only find agreement with our SBDB simulations without massive Plutinos and Twotinos (blue lines in both panels), having a very close match for the 2:1 MMR, and a good match to the initial slope in the 3:2 resonance. Since the real Trans-Neptunian region contains massive objects, it stands to reason that the erosion of the MMRs is faster than what our GP-only simulations or Greenstreet equations imply, suggesting that the MMRs were more heavily populated than the Greenstreet model predicts, at least for the past 1 Gyr (based on a short extrapolation of our equations into the past). This in turn has implications for the age of Pluto's (and Charon's) surface, with our model suggesting younger surfaces than predicted by \citet{Greenstreet15}.

\subsection{Effect of massive Plutinos and Twotinos on the evolution of resonant populations}

\begin{figure}
    \centering
    \includegraphics[width=\linewidth]{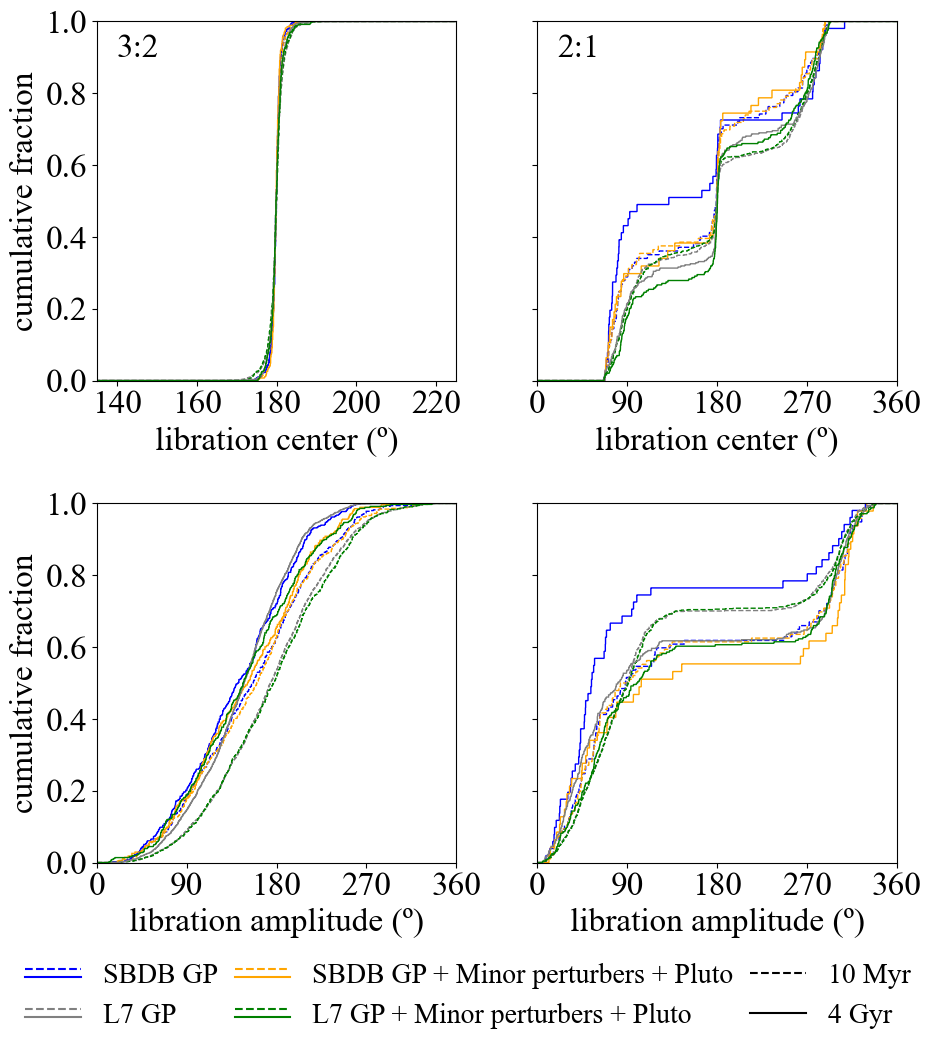}
    \caption{Cumulative fraction of libration amplitude and centers for every population following the same scheme as the one shown in figure \ref{fig:cumm_fractions} with the addition of massive perturbers.}
    \label{fig:cum_massive}
\end{figure}

To further quantify the effect that our sample of massive Plutinos and Twotinos have on the overall evolution of the 3:2 and 2:1 resonant populations, in Figure \ref{fig:cum_massive} we present the cumulative distribution of libration amplitudes and libration centers, comparing the initial and final conditions of the simulations, similar to Figure \ref{fig:cumm_fractions}. Fig. \ref{fig:cum_massive} includes our most complete simulations, i.e., the evolution of the 3:2 population with the ten most massive Plutinos and the evolution of the 2:1 population with the ten most massive Twotinos plus Pluto; both of these scenarios were studied for the SBDB and the L7 model. We also include the curves corresponding to simulations with the giant planets only (the same presented in Fig. \ref{fig:cumm_fractions}), as a point of comparison for the effect of the massive minor perturbers on the secular evolution of the resonant populations.

The results for the 3:2 population (left panels) show that the cumulative distributions are almost unaffected by the presence of additional perturbers, with libration centers (upper left panel) remaining closely aligned with 180 degrees, for both the SBDB and the L7 populations. Libration amplitudes (lower left panels) with only the giant planets will evolve to have a smaller fraction of objects with high libration amplitudes (blue and gray solid lines); however, with the inclusion of massive perturbers, this effect will be reduced, probably because these massive objects can repopulate the high libration orbits by dynamically heating objects with lower libration amplitudes (as shown in green and orange solid lines). It should be noted that the simulation with the smallest evolution is the SBDB model with massive Plutinos included, suggesting that, from all of our simulations, the SBDB population when embedded in a system with (these specific) TNO perturbers is the one that best matches the dynamical system where we embedded them.

In contrast to the 3:2 MMR, there are significant differences present in the 2:1 population, especially for the SBDB simulations after including the ten massive Twotinos and Pluto. To begin with, the fractional population of the symmetric and assymmetric libration centers is not significantly modified from the initial distribution when Pluto and the massive Twotinos are present in the simulations, as can be seen from the yellow lines in the upper right panel of Fig. \ref{fig:cum_massive}. This means that the drastic changes observed in the relative importance of the asymmetric islands in simulations that only include the giant planets are artificial, and the most complete model (especially considering the inclusion of Pluto) results in these fractions not being modified. Thus, we can conclude that either Pluto is essential to protect the originally imprinted asymmetry between the populations trapped in the leading and trailing islands, or such asymmetry is the result of the evolution alongside Pluto. 

Regarding libration amplitudes of the 2:1 population, the effect is similar to the one observed in the 3:2, i.e., for the SBDB, the inclusion of massive perturbers has a drastic effect, incrementing the fraction of particles with large libration amplitudes (solid orange line). Regarding the libration center, the inclusion of massive perturbers in the evolution of the SBDB particles negates the asymmetry created in the GP-only model, finding (again) a very similar distribution at the beginning and at the end of our simulation. On the other hand, for the L7 population, we do not observe significant differences in libration centers nor libration amplitudes, with or without massive Twotinos and Pluto. 

The strong difference between the evolutions of the SBDB particles is likely due to the objects being strongly affected by the removal of the massive TNOs; on the other hand, the weak response present in the L7 model is likely due to the time it needs to find equilibrium with the GPs; perhaps due to the lack of limits imposed in the L7 model, which is generated from random distributions; naturally, these distributions tend to fill all the available phase space. Of course, by statistically reproducing observations, a good portion of the unstable phase space is in practice avoided, but the finer details are not captured. This shows that, despite containing significant biases, the SBDB sample is closer to representing the actual, naturally stable, phase space.

\section{Conclusions} 
\label{sec:conclusions}

This paper studied the long-term dynamical behavior of the 3:2 and 2:1 Neptune mean motion resonant populations in the Kuiper belt. Using theoretical and observational data from the JPL's Small-Body Database and the L7 synthetic model of the Kuiper belt, we performed short-term simulations to characterize the initial populations and long-term simulations to study their evolution, using REBOUND. For each dataset, we ran simulations with different configurations of massive objects besides the Sun. We explored three scenarios for the 3:2 MMR: 1) the 4 GPs of the solar system; 2) the 4 GPs and the 10 most massive Plutinos (including Pluto); and 3) the 4 GPs and Pluto as the only massive perturber. For the 2:1 MMR we explored 4 scenarios: 1) the 4 GPs of the solar system; 2) the 4 GPs and the 10 most massive Twotinos; 3) the 4 GPs and Pluto as a massive perturber (which is in a 4:3 resonance with the 2:1 MMR); and 4) the 4 GPs, the 10 most massive Twotinos, and Pluto.

In section \ref{subsec:long_term}, we conducted an analysis of the final state of the simulations that include only the GPs after 4 Gyr of integration. The sculpting of the orbital parameters tends to converge for the SBDB and the L7 populations, especially in semimajor axis. However, the median inclination of the 3:2 population, as well as the median eccentricity of the 2:1 population, did not really converge, indicating that no amount of evolution will make these distributions consistent among each other. The SBDB sample has observational biases, and whether there are inclination over or underestimations in the L7 model is beyond the scope of this paper. We also note that larger inclinations facilitate the escape from resonance, thus particles in the large-$i$ tail leak faster; this results in the median inclinations of all the populations being smaller at 4 Gyr.

The comparison between the initial and final distribution of the libration amplitudes and libration centers revealed that asymmetries in the leading and trailing populations of the 2:1 population remain unchanged over the evolution of our simulations: L7 objects, which start with a symmetrical distribution, end up with the same rough amount of leading and trailing objects, and the SBDB objects, which start with an uneven distribution among libration centers, ends up with a different, still uneven distribution at the 4 Gyr mark. This suggests that asymmetries in the leading and trailing populations probably arise due to a secondary mechanism at the early stage of the formation of the original populations, most commonly attributed to Neptune's migration at the early stages of the Solar System \citep[e.g.,][and references therein]{Nesvorny18}.

%ççççççç casi listo para incluirse en el parrafo anterior ... \textbf{Given that the SBDB distribution is biased, and that the L7 distribution does not accurately represent the resonances, particularly the 2:1 resonance, for which the modeling is based on only five objects, a more comprehensive study of the evolution of the 2:1 resonant islands, therefore requires a larger and better population. Perhaps by combining \citet{Chen21}'s model with new LSST observations, we could gain more sound insights into the evolution of the resonant island populations.}

By adding the massive perturbers to the simulations, as discussed in Section \ref{subsec:massives}, we explored the effect of these additional perturbations among themselves, as well as over the resonant populations as a whole. Regarding the massive objects, from the ten most massive Plutinos, we found that when only the GPs are considered, all ten objects are resonant for the whole 4 Gyr, while adding mass results in 9 being resonant for 4 Gyr, while the remaining one, \dpn{2003}{UZ}{413}, is resonant for about 1.5 Gyr. For the massive Twotinos, even when only the GPs are included as perturbers, only two massive Twotinos are resonant for the whole 4 Gyr (although a third one is recaptured near the end of the simulation); the other 8 leave the resonance at different times between 20 Myr and 3.2 Gyr. When adding mass to these Twotinos, they become more stable, with six remaining resonant for the 4 Gyr (and one more being recaptured). Yet when adding Pluto as a massive TNO, the set becomes less stable with only four massive Twotinos remaining resonant for the 4 Gyr (and, once more, one more being recaptured).   

The simulations, both with and without the massive TNOs, showed a constant decrease in the size of each resonant population, i.e., a continuous leaking process that can be described remarkably well by an exponential decay (characteristic of stochastic processes) of semi-stable objects, plus a constant offset introduced by an underlying stable population. The fitting parameters of the exponential curves varied greatly for each subset of data. 

The addition of massive bodies into the simulations had a prominent effect on each group, drastically diminishing their stability in some cases, particularly for the 3:2 population. Within the 4 Gyr of our simulations, Pluto is responsible for more than 24\% of the objects escaping the 3:2 population, while only 2\% is attributed to the other nine most massive Plutinos. The ten most massive Twotinos create a difference of 4\% in the number of objects leaving the resonance. The most surprising feature found is the influence of Pluto on the 2:1 population, where 13\% of the escapes are caused by the addition of Pluto alone. The influence of Pluto over the stability of the 2:1 MMR is quite unexpected. It could be attributed to perturbations from the dwarf planet when it lies closer to aphelion, when its distance to objects trapped in 2:1 MMR could be drastically reduced, increasing the interaction probability, but if such were the case, Eris should be as relevant for the 3:2 population stability given its perihelion of $\sim38.4$ au. Preliminary test simulations showed that Eris does not significantly perturb the 3:2 population; thus, a different possibility is that Pluto affects the 2:1 MMR populations through their mutual 4:3 MMR. A more in-depth study of this phenomenon is left for future work (Ram\'irez-Vargas et al. in prep.).

When comparing our results with those of previous authors, we found that simulations of the SBDB without massive perturbers are mostly consistent with extrapolations of the theoretical models \citep[e.g.][]{Greenstreet15}, however, our simulations including massive TNOs result in greater leaking rates.

We found that simulations of the SBDB with the inclusion of massive Plutinos and Twotinos plus Pluto, were the ones which consistently showed least evolution in the cumulative distribution of resonant centers and libration amplitudes; this indicates that the initial distribution was already closest to the shape implied by the perturbers included in our simulations, suggesting that both these ingredients are the ones closest to reality.

The presence of Pluto is sufficient to maintain the observed ratio between the asymmetric islands of the 2:1 MMR. However, further studies are required to determine whether Pluto alone can produce the observed asymmetry, or whether these differences were imprinted during the early evolutionary stages of the Solar System, most likely as a consequence of Neptune’s outward migration. In any case, Pluto’s influence on the dynamical evolution of the outer Solar System is so significant that modern dynamical simulations must include it alongside the giant planets.

\begin{acknowledgments}

We acknowledge the referee, B. Gladman, for a careful reading and report that helped improve this paper. This research was performed using services/resources provided by Grid UNAM, which is a collaborative effort driven by DGTIC and the research institutes of Astronomy, Nuclear Science and Atmosphere Science and Climate Change at UNAM. M.A.M. acknowledges Universidad de Atacama for the DIUDA grant No. 88231R14. A.P.-V. acknowledges the DGAPA-PAPIIT grants IA103224 and IN112526.

\end{acknowledgments}

\begin{contribution}

%%This section gives authors the space to recognize author contributions. The text inside this environment is NOT counted towards the total word quanta. At a minimum, manuscripts are expected to include this text:

MAMG and AP came up with the initial research concept, most of the writing, and most of the editing. SR was responsible for the computer simulations, as well as some writing and editing. APV and CP contributed to reviewing, writing, and editing.

%% But authors are expected to provide more specific details, e.g. 
%%
%%SC was responsible for writing and submitting the manuscript.
%%WWM came up with the initial research concept and edited the manuscript.
%%OTS obtained the funding and edited the manuscript.
%%EBF provided the formal analysis and validation. He also edited the manuscript.
%%GEH Supervised the undergraduates, wrote the software and administers the project github and Zenodo repositories.
%%
%% Authors can use the Contributor Role Taxonomy (CRediT) at
%% for ideas on how write a good statement tailored to their needs.

\end{contribution}


%\facilities{HST(STIS), Swift(XRT and UVOT), AAVSO, CTIO:1.3m, CTIO:1.5m, CXO}

%\software{astropy \citep{2013A&A...558A..33A,2018AJ....156..123A,2022ApJ...935..167A},  
 %         Cloudy \citep{2013RMxAA..49..137F}, 
  %        Source Extractor \citep{1996A&AS..117..393B}
     %     }

\software{This work has made use of the integrator package Rebound \citep{rebound}, and the {\sc Python} modules {\sc Matplotlib} \citep{Hunter07}, and {\sc NumPy} \citep{Harris20}.}


%\appendix

%\section{Appendix information}


\bibliography{dpsbib}{}
\bibliographystyle{aasjournal}



\end{document}